\documentclass[iicol]{sn-jnl}
\usepackage{amsmath}
\usepackage{xcolor}
\usepackage{amssymb}
\usepackage{subcaption}
\newcommand{\sbt}{\,\begin{picture}(-1,1)(-1,-3)\circle*{3}\end{picture}\;\,}

\newcommand{\vect}[1]{\boldsymbol{#1}}
\newcommand{\tens}[1]{\boldsymbol{\mathsf{#1}}}
\newcommand{\Tens}[1]{\mathbb{#1}}
\DeclareMathOperator{\Tr}{Tr}
\newcommand\iWidth{5.5}
\newcommand\LineS{0.39}
\jyear{2022}%

\theoremstyle{thmstyleone}%
\theoremstyle{thmstyletwo}%

\theoremstyle{thmstylethree}%

\begin{document}
\title[Finite strain poro-hyperelasticity]{Finite strain poro-hyperelasticity: An asymptotic multi-scale ALE-FSI approach supported by ANNs}
\author*[1]{\fnm{Hamidreza} \sur{Dehghani}}\email{dehghanihamid2@gmail.com}
\author[1]{\fnm{Andreas} \sur{Zilian}}\email{andreas.zilian@uni.com}
\affil[1]{\orgdiv{Institute of Computational Engineering and Sciences, Department of Engineering, Faculty of Science, Technology and Medicine},  \orgname{University of Luxembourg}, \orgaddress{\street{6 Avenue de la Fonte}, \postcode{4364} \city{Esch-sur-Alzette}, \country{Luxembourg}}}

\abstract{\textcolor{red}{
This contribution introduces and discusses a formulation of poro-hyperelasticity at finite strains. The prediction of the time-dependent response of such media requires consideration of their characteristic multi-scale and multi-physics characteristics.
In the present work this is achieved by formulating a non-dimensionalised fluid-solid interaction problem (FSI) at the pore level using an arbitrary Lagrange-Euler description (ALE). The resulting coupled system of PDEs on the reference configuration are expanded and analysed using the asymptotic homogenisation technique. This approach yields three partially novel systems of PDEs: the macroscopic/effective problem and two supplementary microscale problems (fluid and solid). The latter two provide the microscopic response fields whose average value is required in real-time/online form to determine the macroscale response (a concurrent multi-scale approach).
In order to overcome the computational challenges related to the above multi-scale closure, this work introduces a surrogate approach for replacing the direct numerical simulation by an artificial neural network. This methodology allows for solving finite strain (multi-scale) porohyperelastic problems accurately using direct automated differentiation through the strain energy. Optimal and reliable training data sets are produced from direct numerical simulations of the fully-resolved problem by including a simple real-time output density check for adaptive sampling step refinement. The data-driven approach is complemented by a sensitivity analysis of the RVE response.
The significance of the presented approach for finite strain poro-elasticity/poro-hyperelasticity is shown in the numerical benchmark of a multi-scale confined consolidation problem. Finally, to show the robustness of the method, the system response is dimensionalised using characteristic values of soil and brain mechanics scenarios.
}}

\maketitle

\section{Introduction} \label{intro}
Porohyperelastic materials present multi-scale biphasic problems in which a hyperelastic porous medium interacts with viscous fluid filling and percolating through its pores.
 
 For decades,  the nonlinear multi-scale phenomena including heterogeneous elasticity and  poroelastic/porohyperelastic problems have been under investigation (especially, from the theoretical point of view) \cite{Zohdi2007, Bio, BERRYMAN198597, Berryman2005, Bemer2001}. However, a robust multi-scale multi-physics methodology for solving finite strain poro-hyperelasticity (from theory to general computational mechanics solutions at both scales) was missing since the algorithms are multiplex requiring multiple domains of knowledge/expertise and great computational power. In \cite{Brown2014} the ALE description is used to achieve a linear poroelastic model with infinitesimal pore-scale deformation assumption. This was not solved at the homogenised level although it was simplified by neglecting the effects of macroscopic deformation on RVE mechanical response. In \cite{Miller2021} this problem is approached theoretically using asymptotic homogenisation of the Lagrangian description of the Fluid-Solid Interaction (FSI) at the microscale. However, since the fluid phase can introduce extremely large "deformations", this approach might be suffering from inaccuracy and numerical instability.
Although the FSI problems in the Lagrangian setting have been addressed using advanced techniques (including mesh-moving and mixed-hybrid velocity-based formulations \cite{zilian2009, zilian2011, zilian2010}) the complexity imposed by the multi-scale nature of porohyperelastic problems does not allow for embracing Lagrangian techniques. 

\begin{figure*}
\centering
	\includegraphics[width=  15  cm ]{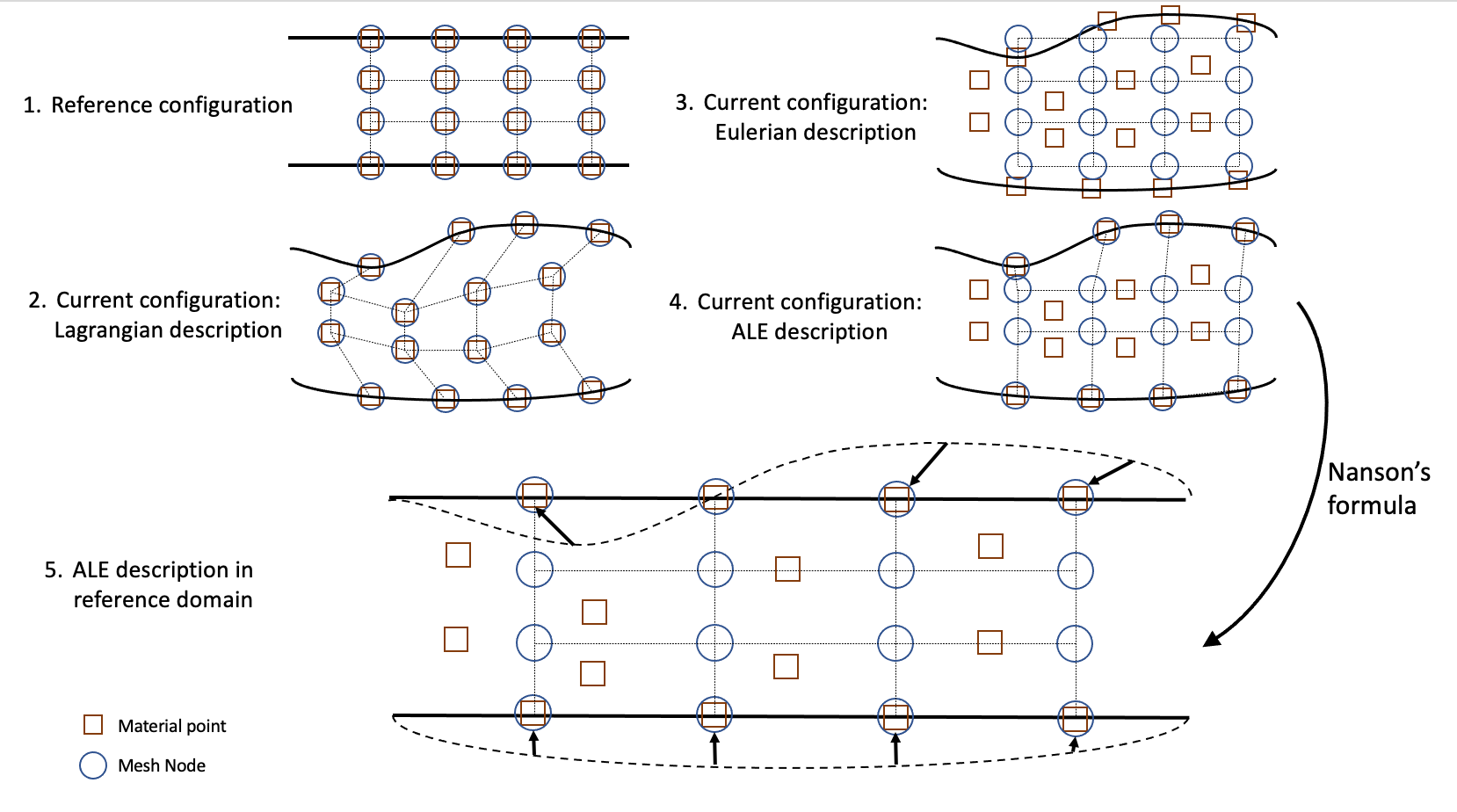}
	\centering
	\caption{\textcolor{red}{Schematic representation of different configurations and descriptions in the fluid domain. The ALE computational mesh follows the motion of solid material points and fluid-solid interface (Lagrangian), while it is fixed (Eulerian) inside the fluid domain. We use Description 5. to reduce the computational cost, since it does not require solving a fluid problem at every quadrature point and increment of the homogenised problem.}}
	\label{Configs_desc}
\end{figure*}

\textcolor{red}{In general, the structure of the multi-scale ALE-FSI approach includes the formulation of the FSI problem at the physical scale (before scales decoupling) using the ALE description shown, schematically, in Figure \ref{Configs_desc}. This is followed by the multi-scale analysis of the achieved equations (ALE-FSI) using asymptotic (two-scales) homogenisation which results in three systems of PDEs for the homogenised (macroscale) problem, solid cell (RVE) problem, and fluid cell problem. The microscale solid response is required at every quadrature point and increment to determine the macroscale response, which is very time-consuming. Thus, we construct an Artificial Neural Network (ANN) as a surrogate for microscale problems delivering the required results in real-time. The mentioned techniques are explained further in the corresponding sections.}

In this study, we employ the ALE method, which is widely embraced for fluid flow with moving boundaries \cite{DUARTE20044819, Donea2017}, multidimensional fluid dynamics problems \cite{HIRT1974227}, FSI problems \cite{Baffico2005}, etc. The ALE method uses an arbitrary computational mesh that could be moved in any prescribed manner (including Lagrangian) or could be held fixed (Eulerian). We employ this approach since the fluid is flowing within a domain that is deforming/moving due to solid deformation. The arbitrary mesh is chosen to move with material (Lagrangian) in the solid skeleton and on the interface while it is held fixed within the fluid domain (except for the fluid-solid interface). The balance of mass and momentum equations of the fluid are also mapped to the \textit{reference domain} (e.g. the fluid domain with undeformed solid phase) using the stress continuity on the interface (using Piola transformation), which is employed to establish the effective balance of linear momentum. This approach provides an economical, accurate, and numerically stable approximation of the model response as it does not directly include fluid displacement or fluid deformation.
Since the problem has multiple length scales the ALE formulation of the FSI problem is non-dimensionalised to avoid dimension-related ambiguities. Furthermore, for the first time, we formulate this type of mathematical multi-scale problem directly using the first-order derivative of a hyperelastic strain energy density function (neo-Hookean) with respect to the deformation gradient tensor (no explicit linearisation of the system) which again results in higher accuracy and numerical stability.

Assuming a sharp length scale separation (between micro and macro scales) and initial local periodicity, the problem can be \textit{regularised}, thus standard \textit{homogenisation} approaches can be adopted \cite{Zohdi2005}. Here, the resulting dimensionless ALE-FSI formulation is analysed using two-scale asymptotic homogenisation techniques \cite{Brown2014, Miller2021, burrigekeller, Pentaasymptotic2017}. This includes differential operator decoupling and power series representation of the relevant fields which leads to the expanded form of the equations. Analysing the latter, for the macroscale, the effective stress, balance of linear momentum, mass conservation, and transformed form of Darcy's law are accompanied by the corresponding microscale hyperelastic solid and fluid RVE/cell systems of PDEs. This is followed by developing the weak formulation and the incremental analysis of the problem via the Finite Element (FE) method.

The general numerical strategy for solving the problem is similar to the FE2 method for composites (see e.g. \cite{SMIT1998181, Schroder2014}) in the sense that for each time increment and each macroscale numerical quadrature the responses of microscopic problems are required. In FE2, the volume average of stress is determined at the microscale (without any constitutive law at the macro level) while, here, we calculate the average microscale displacement gradient tensor using RVE problems at the microscale. The latter is supplied to the macroscale constitutive law to determine the effective stress. This has the advantage of constructing the deformation gradient and Piola transformation tensors directly at the macroscale, allowing us to solve the mass conservation equation and calculate the transformed pore pressure for the determination of the effective stress. This type of online calculation could be computationally very expensive and, in case of larger problems, infeasible. 
However, if we can obtain the average displacement tensor differently there is no need for solving  for the full displacement field as part of the effective system of PDEs. This can be achieved with considerable speed-up by exploiting the predictive power of ANNs in the context of computational mechanics \cite{Rosenblatt58theperceptron, HDAZ2020, HDAZporo2021, DEHGHANI2021398}.

When constructing a suitable ANN to serve as a surrogate for the microscale system of PDEs (a continuous function which is relatively time-consuming to be solved) we consider the macroscale displacement gradient and pore pressure as the inputs while the average microscale displacement gradient tensor is the output. The ANN needs to be \textit{trained} to deliver accurate and reliable outputs. The training procedure consists of tuning the ANNs parameters (weights and biases) by minimising a \textit{cost function}. The latter represents the distance between the ANN outputs and the "exact" values of a sufficient number of samples that are provided in the \textit{training dataset}. In fact, the final value of the minimised cost function (here, obtained using Adam optimiser \cite{kingma2014adam}), shows how accurate the ANN represents the features introduced by the training dataset. However, in order to reach the desired fidelity, one should ensure that the training dataset also reflects the features of the original function. To this end, in this study, a simple real-time output density check algorithm is presented providing an optimal reliable density of the training dataset. 
Furthermore, to demonstrate the microscale response under a variety of possible macroscale displacement gradient components and pore pressure, we perform a sensitivity analysis on RVE problems and study the local phenomena (local concentration of strain energy density) in detail.

The above approach is implemented using the open-source package FEniCS \cite{AlnaesEtal2015} and the open-source Machine Learning package Pytorch \cite{paszke2017automatic} for ANNs. A confined consolidation/compression test is used to verify the numerical implementation and to gain a deeper understanding of the importance and robustness of the present method for finite strain porohyperelastic problems. The results of the numerical examples show that the maximum strain energy density and deformation are observed at the intersection of the pores with values considerably above the average. The deformation predicted by the present method is smaller than the one obtained using the linear poroelastic method (due to the strain stiffening feature of the neo-Hookean model). The steady-state is achieved in a considerably shorter time, and considerable variations in tangent properties in time and space can be observed.
Finally, the non-dimensional variables are dimensionalised using the characteristic values of two scenarios of interest, namely, brain tissue and soil mechanics.

The ALE formulation together with asymptotic/two-scale homogenisation and the final macroscale and microscale governing PDEs are provided in Section \ref{mathDescription}. The weak formulation, incremental analysis and data-driven approach together with sensitivity analysis of the RVE response are carried out in Section \ref{Solution_s}. Section \ref{numericalEX} presents a benchmark finite strain porohyperelastic problem (a consolidation test) using the present method. Finally, Section \ref{Conc} provides the concluding remarks of the present work followed by some potential future directions.

\section{Mathematical description} \label{mathDescription}
Let us assume a poroelastic body in physical (single) scale (domain $\Omega$) which consists of a subdomain of hyperelastic porous skeleton $\Omega_s$ (reference solid domain) and the complementary viscous fluid $\Omega_f$ (reference fluid domain) filling and percolating the pores such that $\Omega = \Omega_s \cup \Omega_f$.
In this section, the mathematical description of the ALE-FSI problem at the pore level is followed by asymptotic homogenisation/multi-scale analysis resulting in the effective governing system of PDEs and the corresponding solid and fluid RVE/cell problems.

\subsection{ALE-FSI governing equations}
The Finite strain solid problem is described using the Lagrangian method in which the observer follows the material movement. The first Piola-Kirchhoff stress (PK1) is employed which is a two-point tensor consistent with the Lagrangian displacement gradient. Furthermore, PK1 is calculated using the deformation gradient tensor which does not involve higher order displacement gradients. This renders it ideal for asymptotic homogenisation.

Since the fluid domain obeys the solid deformation (moving boundary/domain problem) the Eulerian method with the stationary coordinate system (which is usually employed in fluid mechanics) leads to inaccuracies. Due to the latter condition, (fluid flow with moving boundaries) we employ the ALE method which includes a mesh that follows the moving boundaries \cite{DUARTE20044819}. The ALE description is chosen to follow the material/Lagrangian description on the interface and within the solid domain while it uses the Eulerian/spatial description elsewhere in the fluid domain. Thus, one can formulate the fluid problem in the current configuration but within the reference fluid domain by pulling back the interface from the deformed configuration to the reference one using the ALE method.
The fluid-solid coupling is provided by the interface transformation from the deformed configuration to the reference one using Nanson's formula.

 
\subsubsection{Solid}
The spatial variables $\bar{\vect X}$ and $\bar{\vect x}$ refer to reference/undeformed and current/deformed configurations in physical scale (before multi-scale considerations), respectively.
The solid problem is described using the balance of linear momentum
\begin{align}
 \vect 0 &= \nabla_{\bar{\vect X}} \cdot \tens P & \textrm{in} \quad \Omega_s \label{SolBal}
 \intertext{where}
 \tens P &= \frac{\partial \Psi}{\partial \tens F} & \textrm{in} \quad \Omega_s
  \intertext{and}
\tens F &= \nabla_{\bar{\vect X}} \vect u + \tens I & \textrm{in} \quad \Omega_s
\end{align}
are the first Piola-Kirchhoff (PK1) stress and the deformation gradient tensor, respectively. Furthermore, $\Psi$ and $\vect u$ indicate solid strain energy density function and displacement vector, respectively.

\subsubsection{Interface and Fluid}
Neglecting the inertial forces and convective term due to the standard fluid velocity scaling (explained in Section \ref{Sec_nondim}) 
\textcolor{red}{ the ALE form of the fluid problem description reads \cite{DUARTE20044819, Donea2017}}
\begin{align}
\vect 0 &=\nabla_{\bar{\vect x}} \cdot \tens \sigma  & \textrm{in} \quad \Omega_f \label{eqFluidBal}\\
0 &=\nabla_{\bar{\vect x}} \cdot \vect v & \textrm{in} \,\,\,\, \Omega_f. \label{eqFluidMass}
\intertext{The Cauchy fluid stress $\tens \sigma$ is defined as}
\tens \sigma & = -p \tens I + \mu^f \left(\nabla_{\bar{\vect x}} \vect v + \left(\nabla_{\bar{\vect x}} \vect v\right)^T\right) & \textrm{in} \,\,\,\, \Omega_f, \label{eqFluidConsD}
\end{align}
where $\mu^f$, $\vect v$, and $p$ are the fluid dynamic viscosity, velocity and pressure.

In order to provide accurate interface conditions, the transformation of the interface area from deformed/current configuration to reference configuration is necessary. The latter is possible using Nanson's formula as follows
\begin{align}
\vect n da &= \tens G^T \cdot \vect N dA, \label{Nanson}
\intertext{where $\vect n$ and $\vect N$ indicate the normal vectors to the interface surface in deformed and reference configurations, respectively. The second order Piola transformation tensor $\tens G$ is}
\tens G &= J\tens F^{-1}. 
\intertext{Thus, the stress continuity on the interface is established as}
\tens \sigma \tens G^T  \cdot \vect N &= \tens P \cdot \vect N &\textrm{on} \quad \Gamma. \label{inter} 
\intertext{Furthermore, Nanson's formula together with divergence theorem provide a coordinate transformation for arbitrary vector $\vect V$ and tensor $\tens T$}
\nabla_{\bar{\vect x}} \cdot \vect V &= \frac{1}{J} \nabla_{\bar{\vect X}} \cdot (J \tens F^{-1}\vect V) \label{Vect_transform}\\
\nabla_{\bar{\vect x}} \cdot  (\tens T) &= \frac{1}{J}\nabla_{\bar{\vect X}} \cdot (J \tens T \tens F^{-T}) \\
\nabla_{\bar{\vect x}} (\sbt) &= \nabla_{\bar{\vect X}}(\sbt) \tens F^{-1}. \label{gradient_transform}
\intertext{
Thus, Equations \eqref{eqFluidBal} and \eqref{eqFluidMass} can be pulled back to a fixed reference fluid domain as}
\vect 0 &= \nabla_{\bar{\vect X}} \cdot (\tens \sigma \tens G^T) & \textrm{in} \quad \Omega_f \label{eqLFluidBal}\\
0 &= \nabla_{\bar{\vect X}} \cdot (\vect v \tens G^T) & \textrm{in} \,\,\,\, \Omega_f, \label{eqLFluidMass}
\end{align}
to be used in establishing the effective balance of linear momentum.

\subsection{Non-dimensionalisation} \label{Sec_nondim}
\textcolor{red}{The problem at hand involves two or more length scales that vary in an application-specific range (e.g. biomechanics, soil and rock mechanics). Moreover, due to the multi-physics nature of the matter, there is a complex interdependency between the units of measurement. Thus,  the dimensions can become a potential source of ambiguity leading to a misunderstanding of the model parameters and response \cite{Firdaouss1997}. To address this issue, we consider the non-dimensionalisation procedure before proceeding to the multi-scale formulation.} The non-dimensionalisation is similar to the ones in \cite{ HDAZporo2021, growing, pentaBook2019} but extended for ALE formulation and provided here for clarification and the reader's convenience. 
Let us define $\vect X$ and $\vect Y$ denoting, respectively, the formally independent macroscale and microscale reference spatial variables, defined as
\begin{align}
\vect Y {:}&= \frac{\vect X}{\epsilon},
\end{align}
and $\vect x$ and $\vect y$ being the current counterparts.
 $0 <\epsilon = \frac{d}{L} \ll 1$ is the scale separation factor and $d$ and $L$ are, respectively, microscale (RVE) and macroscale representative lengths.
We define two formally independent representative values for fluid dynamic viscosity $\mu_c^f$ and force $f_c$ such that
\begin{gather}
 \vect{X}=L\vect{X}',\quad \vect Y = d \vect Y', \quad \vect{x}=L\vect{x}', \\ \nonumber
 \vect y = d \vect y', \quad \vect f = f_c \vect f', \quad  \mu^f=\mu_c^f \mu'^f, \label{nondim2}
 \end{gather}
where $\vect f$ is (any) force, and $\mu^f$ is the interstitial fluid dynamic viscosity. All other parameters in this study are non-dimensionalised with respect to the mentioned independent characteristic values e.g.
\begin{gather}
\nonumber  \vect u = L \vect u',\quad \vect{v} =\frac{f_cd^2}{L^3\mu_c^f}\vect{v}', \quad \tens P=\frac{f_c}{L^2 }\tens P', \\ \nonumber \quad t = \frac{L^4 \mu_c^f}{f_c d^2} t', \quad \tens{\sigma}=\frac{f_c}{L^2 }\tens{\sigma}' \\ 
\mu^s = \frac{f_c}{2 L^2 (1+\nu)} \mu'^s, \quad \lambda^s = \frac{f_c \nu}{L^2 (1+\nu) (1-2\nu)} \lambda'^s \label{non_dim}
\end{gather}
where $\mu^s$ and $\lambda^s$ are the solid Lam\'e constants (to be used in the solid constitutive law), $\nu$ is the solid Poisson's ratio, and $\sbt'$ indicates the non-dimensional parameter. 

Substituting the non-dimensional fields in the governing equations and \textbf{dropping the prime symbol}, which indicates non-dimensionalised variables (for the sake of simplicity), all the equations will have the same form except for Equation \eqref{eqFluidConsD} which, factorising and removing $\frac{f_c}{L^2}$, takes the form \cite{growing, burrigekeller}
\begin{align}
\tens \sigma &= -p \tens I + \epsilon^2 \mu^f \left(\nabla_{\bar{\vect x}} \vect v + \left(\nabla_{\bar{\vect x}} \vect v\right)^T\right) &\textrm{in} \quad \Omega_f. \label{eqFluidCons}
\end{align}
The fluid convective term and inertial effects are also scaled by $\approx \epsilon^4$ (i.e. $(\vect v - \dot{\vect u}) \cdot \nabla_{\bar{\vect x}} \vect v =  \frac{f_c d^4}{L^5(\mu_c^f)^2} (\vect v' - \dot{\vect u'}) \cdot \nabla_{\vect{\bar x'}} \vect v'$ and $\frac{ {\rm d} \vect v}{{\rm d} t} = \frac{f_c d^4}{L^5(\mu_c^f)^2} \frac{{\rm d} \vect v'}{{\rm d} t'}$) ($\mathcal O(\epsilon^4)$) and neglected from the governing equations.

\subsection{Multi-scale expansion} \label{SecExpansion}
Having achieved the dimensionless ALE-FSI governing equations, we can now apply asymptotic two-scale analysis which starts from the multi-scale expansion of the relevant fields using the following steps.
The differential operator is decoupled via
\begin{align}
\nabla &\rightarrow \nabla_{\vect X} + \frac{1}{\epsilon} \nabla_{\vect Y} \label{diff_op}
\intertext{and the fields ($\psi$) are represented via power series}
\psi_\epsilon(\vect X,\vect Y) &= \sum_{l=0}^n \psi^{(l)}(\vect X,\vect Y) \epsilon^l. \label{power_series}
\intertext{\textcolor{red}{We use Equation \eqref{diff_op}  and Equation \eqref{power_series} to achieve the expanded form of all relevant fields, which is provided in Appendix \ref{Expansion}.} The integral average is defined as}
 \left\langle \psi \right\rangle_k &= \frac{1}{\lvert\Omega\rvert}\int_{\Omega_k}\psi(\vect{X},\vect{Y})\,\textrm{d}\vect{Y},\label{intavg_cell} 
\intertext{while the average can be obtained via}
\bar{\psi}_k &= \frac{\left\langle \psi \right\rangle_k \lvert\Omega\rvert}{\lvert\Omega_k\rvert} \quad k=f,s. \label{intavg_domain}
\end{align}
The integral average and average are applied to the parameters and fields with $\vect{y}$-dependency to be employed at the homogenised level.
We carry out the multi-scale expansion of the governing equations in the following.

\subsubsection{\textcolor{red}{Expansion of} governing equations}
Multi-scale expansion of the governing equations is provided here and will be served as the basis for upscaling and to achieve the RVE problems.
 Equation \eqref{SolBal} could be written as

 \begin{align}
\vect 0 &= (\nabla_{\vect X} + \frac{1}{\epsilon} \nabla_{\vect Y}) \cdot (\tens P^{(0)} + \epsilon \tens P^{(1)}) & \textrm{in} \quad \Omega_s. \label{EQ12}
\end{align}
Throughout this study, the coefficients of $\epsilon^{-1}$, $\epsilon^0$, and $\epsilon^1$ are referred to as $\mathcal C(\epsilon^{-1})$, $\mathcal C (1)$, and $\mathcal C (\epsilon^1)$, respectively.
Equation \eqref{EQ12} in $\mathcal C(1)$, $\mathcal C(\epsilon)$ and $\mathcal C(\epsilon^{-1})$ reads, respectively,
\begin{align}
\vect 0 &= \nabla_{\vect X} \cdot \tens P^{(0)} +  \nabla_{\vect Y} \cdot \tens P^{(1)} & \hspace{0.8cm} \textrm{in} \quad \Omega_s \label{contSol}\\
\vect 0 &= \nabla_{\vect X} \cdot \tens P^{(1)} & \textrm{in} \quad \Omega_s\\
\vect 0 &= \nabla_{\vect Y} \cdot \tens P^{(0)} & \textrm{in} \,\,\,\, \Omega_s.\label{DivP0}
\end{align}
\normalsize
Expanding Equation \eqref{eqFluidCons} results in
\begin{align}
\tens \sigma^{(0)} &= - p^{(0)} \tens I \hspace{3cm} \textrm{in} \quad \Omega_f \label{hydro0}\\ 
\tens \sigma^{(1)}  &= - p^{(1)} \tens I + \mu^f \left(\nabla_{\vect y} \vect v^{(0)} + (\nabla_{\vect y} \vect v^{(0)})^T \right). \label{hydro1} 
\end{align}
The expansion of Equation \eqref{eqFluidBal} renders
\begin{align}
\vect 0 &= \nabla_{\vect x} \cdot \tens \sigma^{(0)} + \nabla_{\vect y} \cdot \tens \sigma^{(1)} & \textrm{in} \quad \Omega_f \label{eqFluidBal0}\\
\vect 0 &= \nabla_{\vect y} \cdot \tens \sigma^{(0)}  & \textrm{in} \quad \Omega_f  \label{eqFluidBal-1}\\
\vect 0 &= \nabla_{\vect x} \cdot \tens \sigma^{(1)}  & \textrm{in} \,\,\,\, \Omega_f.  \label{eqFluidBal1}
\intertext{
Similarly, the mass conservation Equation \eqref{eqFluidMass} takes the form}
\vect 0 &= \nabla_{\vect x} \cdot \vect v^{(0)} + \nabla_{\vect y} \cdot \vect v^{(1)} & \textrm{in} \quad \Omega_f  \label{eqFluidMass0}\\
\vect 0 &= \nabla_{\vect y} \cdot \vect v^{(0)} & \textrm{in} \quad \Omega_f  \label{eqFluidMass-1} \\
\vect 0 &= \nabla_{\vect x} \cdot \vect v^{(1)} & \textrm{in} \,\,\,\, \Omega_f.  \label{eqFluidMass1}
\end{align}
Substituting Equation \eqref{hydro0} into \eqref{eqFluidBal-1} highlights that the pore pressure is constant with respect to the microscale space ($p^{(0)} = p^{(0)}(\vect X, t)$) since $\nabla_{\vect y} p^{(0)} = \vect 0$.

 From multi-scale expansion of Equation \eqref{inter} in $\mathcal C(1)$, $\mathcal C(\epsilon)$ we reach
 \begin{align}
 \tens P^{(0)} \cdot \vect N &= -p^{(0)} \tens G^{(0)T} \cdot \vect N &\textrm{on} \quad \Gamma \label{stress0inter}\\
 \tens P^{(1)} \cdot \vect N &= (\tens \sigma^{(1)}\tens G^{(0)T} -p^{(0)} \tens G^{(1)T} )\cdot \vect N &\textrm{on} \,\,\,\,\Gamma. \label{stress1inter} 
\end{align}
 \textcolor{red}{Having the expanded form of the governing equations, we can proceed to the upscaling process.}

\subsection{Upscaling}
In this section, the PDEs governing the homogenised domain ($\Omega_h$) are derived from the expanded governing equations. 
This is followed by establishing suitable cell problems whose solutions provide the required parameters of the homogenised system of PDEs.
\subsubsection{Balance of linear momentum}
From the sum of Equations \eqref{SolBal} and \eqref{eqLFluidBal} we have
 \begin{align}
\vect 0 &= \int_{\Omega_s} \nabla_{\bar{\vect X}} \cdot \tens P  {\rm d}V_s+ \int_{\Omega_f} \nabla_{\bar{\vect X}} \cdot \tens \sigma \tens G^T {\rm d}V_f
\intertext{
which, after the expansion, takes the form \vspace{0.3cm}} \nonumber
\vect 0 &= \int_{\Omega_s} (\nabla_{\vect X} \cdot \tens P^{(0)}  + \nabla_{\vect Y} \cdot \tens P^{(1)}) {\rm d}V_s + \\ \nonumber
&\hspace{0.6cm} \int_{\Omega_f}\big( \nabla_{\vect X} \cdot \tens \sigma^{(0)} \tens G^{(0)T} +\\ 
& \hspace{1.2cm} \nabla_{\vect Y} \cdot (\tens \sigma^{(1)} \tens G^{(0)T}+ \tens \sigma^{(0)} \tens G^{(1)T})\big) {\rm d}V_f.
\end{align}
Applying the divergence theorem (considering the direction of the normal vector) we reach
\begin{align}
 \nonumber
\vect 0 &=  \int_{\Omega_s} \nabla_{\vect X} \cdot \tens P^{(0)} dV_s + \int_{\Gamma} \tens P^{(1)} \cdot \vect N dS_{\Gamma}  + \\ \nonumber 
&\hspace{0.6cm} \int_{\Omega_f} \nabla_{\vect X} \cdot \tens \sigma^{(0)} \tens G^{(0)T} {\rm d}V_f - \\ 
&\hspace{0.9cm} \int_{\Gamma} (\tens \sigma^{(1)} \tens G^{(0)T}+ \tens \sigma^{(0)} \tens G^{(1)T}) \cdot \vect N {\rm d}S_{\Gamma}.
\intertext{
Considering Equations \eqref{hydro0} and \eqref{stress1inter} we conclude $(\tens \sigma^{(1)} \tens G^{(0)T}+ \tens \sigma^{(0)} \tens G^{(1)T}) \cdot \vect N =  \tens P^{(1)} \cdot \vect N$, thus}
\vect 0 &=  \int_{\Omega_s} \nabla_{\vect X} \cdot \tens P^{(0)} {\rm d}V_s +  \int_{\Omega_f} \nabla_{\vect X} \cdot \tens \sigma^{(0)} \tens G^{(0)T} {\rm d}V_f 
 \end{align}
where $V_s$ and $V_f$ indicate, respectively, volume of solid and fluid phases ($\lvert\Omega_s\rvert$ and $\lvert\Omega_f\rvert$) in the reference configuration.

Considering that $p^{(0)}$ and $\nabla_{\vect X} \vect u^{(0)}$ are spatially constant in microscale space ($\vect y$) and averaging the micro-dependent parts of the equations i.e.
\begin{align}
\int_{\Omega_f} \nabla_{\vect X} \cdot (\tens \sigma^{(0)} {\tens G}^{(0)T}) {\rm d}V_f &=  \nabla_{\vect X} \cdot  (p^{(0)} \langle \bar{\tens G}^{(0)T} \rangle_f) \\
\int_{\Omega_s} \nabla_{\vect X} \cdot \tens P^{(0)} {\rm d}V_s &=  \nabla_{\vect X} \cdot \left\langle \bar{\tens P}^{(0)} \right\rangle_s,
\end{align}
\textbf{the macroscale balance of linear momentum} is
\begin{align}
  \vect 0 &=  - \nabla_{\vect X} \cdot \tens P_E & \textrm{in} \quad \Omega_h\, \label{Equi_macro}
\intertext{where $\Omega_h$ indicates the homogenised domain and} 
\tens P_E &= \left\langle \bar{\tens P}^{(0)} \right\rangle_s - p^{(0)} \left\langle \bar{\tens G}^{(0)T}  \right\rangle_f & \textrm{in} \quad \Omega_h. \label{effectiveP}
\intertext{
We take the same approach as in \cite{Miller2021}, where the zeroth order of expansion of Equation \eqref{SolBal} takes the form}
\tens P^{(0)}&= \frac{\partial \Psi^{(0)}}{\partial \tens F^{(0)}}, \label{P0non}
\intertext{thus \textbf{the constitutive law for solid stress} reads}
\left\langle \bar{\tens P}^{(0)} \right\rangle_s&= V_s \frac{\partial \Psi^{(0)}}{\partial \bar{\tens F}^{(0)}} & \textrm{in} \quad \Omega_h, \label{AverageP0}
\intertext{where, using Equations \eqref{intavg_cell} and \eqref{intavg_domain},}
\bar{\tens F}^{(0)} &=  \nabla_{\vect X} \vect u^{(0)} + \overline{\nabla_{\vect Y} \vect u}^{(1)} + \tens I. \label{barF}
\intertext{and}
\bar{\tens G}^{(0)} &= \det{\bar{\tens F}^{(0)}} (\bar{\tens F}^{(0)})^{-1}
\end{align}

\vspace{0.1cm}
\subsubsection{Mass conservation and Darcy's law}

Expanding Equation \eqref{eqLFluidMass} reads
\begin{align}
\nonumber
0 &= \int_{\Omega_f}\Big( \nabla_{\vect X} \cdot (\tens G^{(0)} \vect v^{(0)}) + \\
&\hspace{1.5cm} \nabla_{\vect Y} \cdot (\tens G^{(0)} \vect v^{(1)} + \tens G^{(1)} \vect v^{(0)})\Big) {\rm d}V_f, \label{mass01}
\intertext{
which, applying the divergence theorem and considering local periodicity in the reference configuration, could be rewritten as}\nonumber
0 &= \int_{\Omega_f} \nabla_{\vect X} \cdot (\tens G^{(0)} \vect v^{(0)}) {\rm d}V_f  - \\
&\hspace{0.6cm}  \int_{\Gamma} (\tens G^{(0)} \vect v^{(1)}  +  \tens G^{(1)} \vect v^{(0)} )\cdot \vect N {\rm d}\Gamma,
 \intertext{Considering the compatibility condition on the interface and assuming no-slip interface condition ($\vect v= \dot{\vect u}$ on $\Gamma$) we can rewrite Equation \eqref{mass01} as} \nonumber
0 &= \int_{\Omega_f} \nabla_{\vect X} \cdot (\tens G^{(0)} \vect v^{(0)}) {\rm d}V_f  - \\ 
&\hspace{0.6cm}  \int_{\Gamma} (\tens G^{(0)} \dot{\vect u}^{(1)}  +  \tens G^{(1)} \dot{\vect u}^{(0)} )\cdot \vect N {\rm d}\Gamma,
\end{align}
 which, again, using the divergence theorem takes the form
\vfill\null
 \begin{align}
  \nonumber
 0 &=  \int_{\Omega_f} \nabla_{\vect X} \cdot  {\tens G}^{(0)}  \vect v^{(0)} {\rm d}V_f- \\
 &\hspace{0.6cm} \int_{\Omega_s} \nabla_{\vect Y} \cdot ( {\tens G}^{(0)} \dot{ \vect u}^{(1)} + \tens G^{(1)} \dot{ \vect u}^{(0)}) {\rm d}V_s. \label{cintinuity1} 
\intertext{
Using Equation \eqref{intavg_cell}, Equation \eqref{cintinuity1} can be written as}\nonumber
0 &=  \nabla_{\vect X} \cdot \langle {\tens G}^{(0)} \vect v^{(0)}\rangle_f -  \langle  \nabla_{\vect Y} \cdot  ({\tens G}^{(0)}\dot{ \vect u}^{(1)}) \rangle_s -  \\
&\hspace{0.6cm} \langle \nabla_{\vect Y} \cdot (\tens G^{(1)} \dot{ \vect u}^{(0)})\rangle_s, \label{cintinuity2} 
\intertext{
which can be rewritten as}
 \nonumber 
0 &= \nabla_{\vect X} \cdot \langle {\tens G}^{(0)}  \vect w\rangle_f + \langle\nabla_{\vect X} \cdot  {\tens G}^{(0)}  \dot{\vect u}^{(0)}\rangle_f \\ 
 &\hspace{0.6cm} -\langle  \nabla_{\vect Y} \cdot  ({\tens G}^{(0)}\dot{ \vect u}^{(1)}) \rangle_s -   \langle \nabla_{\vect Y} \cdot (\tens G^{(1)} \dot{ \vect u}^{(0)})\rangle_s, \label{mass1}
 \intertext{
where}
 \langle& {\tens G}^{(0)} \vect w \rangle_f = \langle {\tens G}^{(0)} \vect v^{(0)} \rangle_f - \langle {\tens G}^{(0)} \dot{\vect u}^{(0)} \rangle_f, \label{ansatzv0} 
\end{align}
yet, this equation should be further processed to be employed in the macroscale system of equations.

Using the divergence identity $\nabla \cdot (\tens A \vect v) = 
\vect v \cdot \nabla \cdot \tens A^T + \tens A^T {:} \nabla \vect v$ for arbitrary tensor $\tens A$ and vector $\vect v$, the last two terms of Equation \eqref{mass1} can be written as
\begin{align}
\nonumber
 &\langle  \nabla_{\vect Y} \cdot  ({\tens G}^{(0)}\dot{ \vect u}^{(1)}) \rangle_s  +  \langle \nabla_{\vect Y} \cdot (\tens G^{(1)} \dot{ \vect u}^{(0)})\rangle_s = \\  \nonumber
 &\hspace{0.3cm}\langle \dot{ \vect u}^{(1)} \cdot \nabla_{\vect Y} \cdot {\tens G}^{(0)T} \rangle_s +  \langle {\tens G}^{(0)T} {:} \nabla_{\vect Y} \dot{\vect u}^{(1)} \rangle_s + \\
&\hspace{0.6cm} \langle \dot{ \vect u}^{(0)} \cdot \nabla_{\vect Y} \cdot \tens G^{(1)T} \rangle_s + \langle \tens G^{(1)T} {:} \nabla_{\vect Y} \dot{ \vect u}^{(0)} \rangle_s
\intertext{
considering $\nabla_{\vect Y} \dot{ \vect u}^{(0)} = 0$ and Equation \eqref{DivYG0}, the first and the last terms of the right hand side are zero. Furthermore, applying the divergence theorem twice, we have $ \dot{ \vect u}^{(0)} \cdot \langle \nabla_{\vect Y} \cdot \tens G^{(1)T} \rangle_s   = -  \dot{ \vect u}^{(0)} \cdot \langle \nabla_{\vect Y} \cdot \tens G^{(1)T} \rangle_f$ thus, using Equation \eqref{DivXG0}, $ \dot{ \vect u}^{(0)} \cdot \langle \nabla_{\vect Y} \cdot \tens G^{(1)T} \rangle_s =  \dot{ \vect u}^{(0)} \cdot \langle \nabla_{\vect X} \cdot \tens G^{(0)T} \rangle_f$. Consequently,}
\nonumber
& \langle  \nabla_{\vect Y} \cdot  ({\tens G}^{(0)}\dot{ \vect u}^{(1)}) \rangle_s +  \langle \nabla_{\vect Y} \cdot (\tens G^{(1)} \dot{ \vect u}^{(0)})\rangle_s  = \\ \nonumber
&\hspace{0.3cm} \langle {\tens G}^{(0)T} {:} \nabla_{\vect Y} \dot{ \vect u}^{(1)} \rangle_s  +  \langle \nabla_{\vect X} \cdot {\tens G}^{(0)} \dot{ \vect u}^{(0)} \rangle_f - \\ 
& \hspace{0.6cm}\langle{\tens G}^{(0)T} {:} \nabla_{\vect X} \dot{ \vect u}^{(0)}\rangle_f \label{ABC}
\end{align}

Finally, substituting Equation \eqref{ABC} into Equation \eqref{mass1} and using the averaged $\vect y$-dependent quantities, \textbf{the effective mass conservation} takes the form
\begin{align} 
\nonumber
0 &= - \nabla_{\vect X} \cdot \langle \bar{\tens G}^{(0)}  \vect w \rangle_f - \langle\bar{\tens G}^{(0)T} {:} \nabla_{\vect X} \dot{ \vect u}^{(0)}\rangle_f + \\ 
& \hspace{0.6cm} \langle \bar{\tens G}^{(0)T} {:} \overline{\nabla_{\vect Y} \dot{ \vect u}}^{(1)} \rangle_s,  \label{Eqmass2} 
\end{align}
which, at infinitesimal strains, holds the analogy with the infinitesimal poroelasticity in \cite{Hdehghani}. 
Darcy's law \cite{Darcy1856} is adopted as \textbf{the standard Ansatz/constitutive law for the effective relative fluid velocity}
\begin{align}
\langle \vect w \rangle_f &= - \tens K \nabla_{\vect x}p^{(0)}. \label{Darcy}
\intertext{
Thus,}
\langle \bar{\tens G}^{(0)} \vect w \rangle_f &= -\bar{\tens G}^{(0)}  \tens K  (\bar{\tens F}^{(0)})^{-T} \nabla_{\vect X}p^{(0)}. \label{Darcy_ALE}
\end{align}

The term $\nabla_{\vect Y} \dot{ \vect u}^{(1)}$ and hydraulic conductivity $\tens K$ are to be determined, respectively, using the RVE problems in solid and fluid domains (solid cell problem and fluid cell problem), which are provided in Section \ref{cellProblems}. We highlight that the term $\nabla_{\vect Y} \dot{ \vect u}^{(1)}$, implicitly, provides the fluid-solid coupling term. 

\subsection{RVE problems} \label{cellProblems}
In this section, the systems of PDEs to be solved in RVE domains to determine the micro-driven parameters (i.e. $\nabla_{\vect Y} { \vect u}^{(1)}$ and hydraulic conductivity $\tens K$) are provided.

\subsubsection{Fluid phase}
Since the fluid is assumed to be incompressible and Newtonian flowing at sufficiently small velocities making the inertial convective effects negligible, it is a linear problem. Thus, instead of solving the fluid problem at each pressure gradient, we calculate its tangent which provides the hydraulic conductivity for Darcy's law as in \cite{burrigekeller, growing, Hdehghani, HdehghaniThesis}.
Substituting Equations \eqref{hydro0} and \eqref{hydro1} into \eqref{eqFluidBal0}
\begin{align} 
\nonumber
\vect 0   &= -\nabla_{\vect x} \cdot (p^{(0)} \tens I) -  \nabla_{\vect y} \cdot (p^{(1)} \tens I) + \\
&\hspace{0.6cm} \mu^f \nabla_{\vect y} \cdot (\nabla_{\vect y} \vect v^{(0)} + (\nabla_{\vect y} \vect v^{(0)})^T).
\end{align}
Considering the RVE geometry, local periodicity, no-slip interface condition, Equations \eqref{eqFluidMass-1} and \eqref{ansatzv0} 
\begin{align}
 \vect 0 &=  \mu^f \nabla_{\vect y}^2 \vect w -  \nabla_{\vect y} p^{(1)} -\nabla_{\vect x} p^{(0)}  &\textrm{in} \quad \Omega_f\\
 0 &= \nabla_{\vect y} \cdot \vect w &\textrm{in} \quad \Omega_f\\
 \vect 0 &= \vect w  &\textrm{on} \quad \Gamma.
\intertext{
We consider the following Ansatz to determine the hydraulic conductivity}
p^{(1)} &=  - \vect P_h \nabla_{\vect x} p^{(0)} \\
\vect w &= -\tilde{\tens K}^T\nabla_{\vect x} p^{(0)}
\intertext{
which results in }
 \vect 0 &= \mu^f \nabla_{\vect y}^2 \tilde{\tens K} -  \nabla_{\vect y} \vect P_h + \tens I  &\textrm{in} \quad \Omega_f \label{RVE_F1}\\
 \vect 0 &= \nabla_{\vect y} \cdot \tilde{\tens K} &\textrm{in} \quad \Omega_f\\
 \vect 0 &= \tilde{\tens K} &\textrm{on} \quad \Gamma \label{RVE_F3},
\intertext{
where the hydraulic conductivity of Darcy's law could be calculated via }
\langle\tilde{\tens K}\rangle_f &= \tens K
\intertext{with the uniqueness conditions}
\vect 0 &= \langle \vect P_h \rangle_f.
\end{align}
The solution of the fluid RVE problem can be obtained by concatenating the results of a componentwise analysis as in \cite{Hdehghani}.

\subsubsection{Solid phase}
Equation \eqref{DivP0} and \eqref{stress0inter}, together with periodic boundary condition, construct a solid RVE problem as follows
\begin{align}
\vect0 &= \nabla_{\vect Y} \cdot \tens P^{(0)} & \textrm{in} \quad \Omega_s \label{Solidcell} \\
\vect0 &= \left(\tens P^{(0)} + p^{(0)} {\tens G}^{(0)T}\right) \cdot \vect N , & \textrm{on} \quad \Gamma \label{SolidcellBC}
\intertext{
with the constitutive law Equation \eqref{P0non}, repeated here for the reader's convenience,}
\tens P^{(0)}&= \frac{\partial \Psi^{(0)}}{\partial \tens F^{(0)}}, \label{P0non1}
\intertext{and}
\tens F^{(0)} &= \nabla_{\vect X} \vect u^{(0)} + \nabla_{\vect Y} \vect u^{(1)} + \tens I. 
\end{align}
At the microscale level (RVE problems), the macroscopic variables $p^{(0)}$ and $\nabla_{\vect X} \vect u^{(0)}$ are knowns to impose a condition under which the microscopic response $\nabla_{\vect Y} \vect u^{(1)}$ is to be determined by solving the RVE problem in the solid phase.

\section{Incremental weak form supported by ANNs} \label{Solution_s}
In this section, we provide the formulation and details required for solving the sets of governing equations numerically. Basically, the macroscale problem is solved numerically via FEM which requires the weak form of the corresponding equations. Furthermore, due to fluid-structure interactions (multi-physics), the response is path-dependent which should be solved incrementally. The microscopic response is required to solve the macroscopic equations due to the multi-scale nature of the problem. This is included via the term $\nabla_{\vect Y} { \vect u}^{(1)}$  and hydraulic conductivity in the effective governing equations (see Equation \eqref{barF}, \eqref{Eqmass2} and \eqref{Darcy_ALE}). The latter is a one-time calculation per initial properties (since the fluid model is linear) while the former ($\nabla_{\vect Y} { \vect u}^{(1)}$) is to be determined per initial properties and within each increment for each finite element since the response is nonlinearly correlated to the macroscopic inputs ($\nabla_{\vect X} {\vect u}^{(0)}$ and $p^{(0)}$).

Such a high number of direct numerical simulations (DNS) of the microscopic RVE problems for one macroscopic problem renders it computationally unaffordable. This is compounded by the need for input-output tangents in the macroscopic iterative solver (e.g. Newton-Raphson). This problem is addressed by employing ANNs trained with numerical results of  microscopic RVE problems in fluid and solid phases via DNS. 

In this section, firstly, the procedures of solving RVE problems via FEM are explained and some RVE numerical tests are provided, secondly,  the data-driven approach (as a surrogate for the RVE problems) is introduced and, finally, the solving strategy for the macroscale problem is presented.

\subsection{RVE fluid problems} 
 Following the componentwise analysis of the system of Equations \eqref{RVE_F1}-\eqref{RVE_F3}, the final system of equations to be solved in the fluid domain reads

\begin{align}
\mu^f \nabla^2_{\vect{y}}\vect{\tilde{K}}_i - \nabla_{\vect{y}}\tilde{p}_i + \vect{e}_i = \vect 0 \quad\textrm{in}\,\,\Omega_f \label{fluidequilibrium} \\
\nabla_{\vect{y}} \cdot \vect{\tilde{K}}_i = \vect 0 \quad\textrm{in}\,\,\Omega_f  \\
\vect{\tilde{K}}_i= \vect 0 \quad\textrm{on}\,\,\Gamma \label{fluidbc},
\end{align}
for every $i = 1, 2, 3$, constructing three Stoke's problems each with a unit body force in the direction of $i$-$th$ unit vector of the chosen system of coordinates (i.e. $\vect{e}_i$). Consequently, the tensor $\tens K = \langle K_{ji} \rangle_f$ can be constructed by concatenating the vectors $\langle\tilde{\vect K}_i\rangle_f$.

Finally, the weak form of Equations \eqref{fluidequilibrium}-\eqref{fluidbc} reads 
\begin{align} \nonumber
&\int_{\Omega_f} \mu^f \nabla \tilde{\vect{K}}_i {:} \nabla \delta \vect v {\rm d} V_f  - \int_{\Omega_f} \tilde{p}_i  (\nabla \cdot \delta \vect v)  {\rm d} V_f  \\
& \hspace{0.3cm} +\int_{\Omega_f} \vect{e}_i  \cdot \delta \vect v {\rm d} V_f  =0 \label{weak1}\\
&\int_{\Omega_f} (\nabla \cdot \vect{v}) \delta q {\rm d} V_f  =0 \label{weakinc}
\end{align}
where, $\delta \vect v$ and $\delta q$ are arbitrary test functions. We highlight that, since the RVE fluid problem is linear, incremental analysis is not required. \textcolor{red}{To accurately solve the fluid problems, the corresponding domain is discretised using 57273 tetrahedral elements.}

\subsection{RVE solid problem}
The RVE solid problem described by Equations \eqref{Solidcell} is a non-symmetric problem with a relatively complex geometry of RVE. 
Solving the problem incrementally (applying BCs in small increments) improves the numerical stability, allowing us to consider larger macroscale deformations and porepressure and providing a dense dataset for ANN training. The incremental weak formulation reads
\begin{align} \nonumber
0 &= \int_{\Omega_s} \Big( (\frac{\partial \Psi^{(0)}}{\partial((\nabla_{\vect X} \vect u^{(0)})_{t+\Delta t} + \nabla_{\vect Y} \vect u^{(1)}_{t+\Delta t} + \tens I )})- \\ \nonumber
& \hspace{0.6cm}   (\frac{\partial \Psi^{(0)}}{\partial((\nabla_{\vect X} \vect u^{(0)})_t + \nabla_{\vect Y} \vect u^{(1)}_t + \tens I )}) \Big)  {:} \nabla_{\vect Y} \delta \vect u^{(1)} {\rm d}V_s \\
& \hspace{1cm} + \int_{\Gamma} (p^{(0)}_{t+\Delta t} \tens G^{(0)}_{t+\Delta t} - p^{(0)}_t \tens G^{(0)}_t)\cdot \vect N \delta \vect u^{(1)} {\rm d} S_{\Gamma} \label{SolidRVEweak}
\end{align}
Note that $\nabla_{\vect X} \vect u^{(0)}$ and $p^{(0)}$ are $\vect y$-constants given by macroscale problem that are imposed incrementally. $\delta \vect u^{(1)}$ and $\vect u^{(1)}$ are, respectively, the test function and the function (solution) of this problem. The output to be provided for macroscale problem is $\langle\nabla_{\vect Y} \vect u^{(1)}\rangle_s$. Furthermore, we assume the compressible neo-Hookean model as the leading order strain energy density function as follows
\begin{align}
\Psi^{(0)} &= \frac{\mu^s}{2}(I^{(0)}_1 - 3) - \mu^s \ln(J^{(0)}) + \frac{\lambda^s}{2} \ln(J^{(0)})^2, \label{Neo_Hookean}
\end{align}
where $J^{(0)}$ is defined via Equation \eqref{EQjzero} and $I_1^{(0)} = \Tr (\bar{\tens F}^{(0)T} \bar{\tens F}^{(0)})$. The constants $\mu^s$ and $\lambda^s$ are the material parameters (the Lam\'e constants). The leading order strain energy density function (Equation \eqref{Neo_Hookean}) can be non-dimensionalised using fields in \eqref{non_dim} as
\begin{align}
\nonumber
\frac{f_c}{L^2}\Psi'^{(0)} &= \frac{f_c}{4 L^2 (1+\nu)}(I^{(0)}_1 - 3) - \\  \nonumber 
&\hspace{0.6cm} \frac{f_c}{2 L^2 (1+\nu)}\ln(J^{(0)}) + \\  
&\hspace{0.9cm} \frac{f_c \nu}{2L^2 (1+\nu) (1-2\nu)}\ln(J^{(0)})^2
\end{align}
which, dropping the prime of variables (for simplicity and consistency with the rest of the formulation), takes the form
\begin{align}
\Psi^{(0)} &=  \frac{(I^{(0)}_1 - 3)}{4(1+\nu)} - \frac{\ln(J^{(0)})}{2 (1+\nu)} + \frac{\nu \ln(J^{(0)})^2}{2 (1+\nu) (1-2\nu)} \label{Psi_nonDim}
\end{align}
We highlight that although the unit of strain energy is $f_c L$ the strain energy density (strain energy per unit volume) has the same unit of stress ($\frac{f_c}{L^2}$).

\paragraph{Local periodicity}
Local periodicity is required within each finite element of the macroscale model at the Reference Lagrangian configuration. It allows us to exploit the potential of asymptotic homogenisation at each element. Since the RVEs in a neighbourhood within each finite element receive the same $\vect y$-constant macroscale inputs ($\nabla_{\vect X} \vect u^{(0)}$ and $p^{(0)}$) the initial local periodicity is maintained throughout the problem.

\begin{figure*}
\centering
	\begin{subfigure}{\iWidth cm}
	\includegraphics[width=  \iWidth cm ]{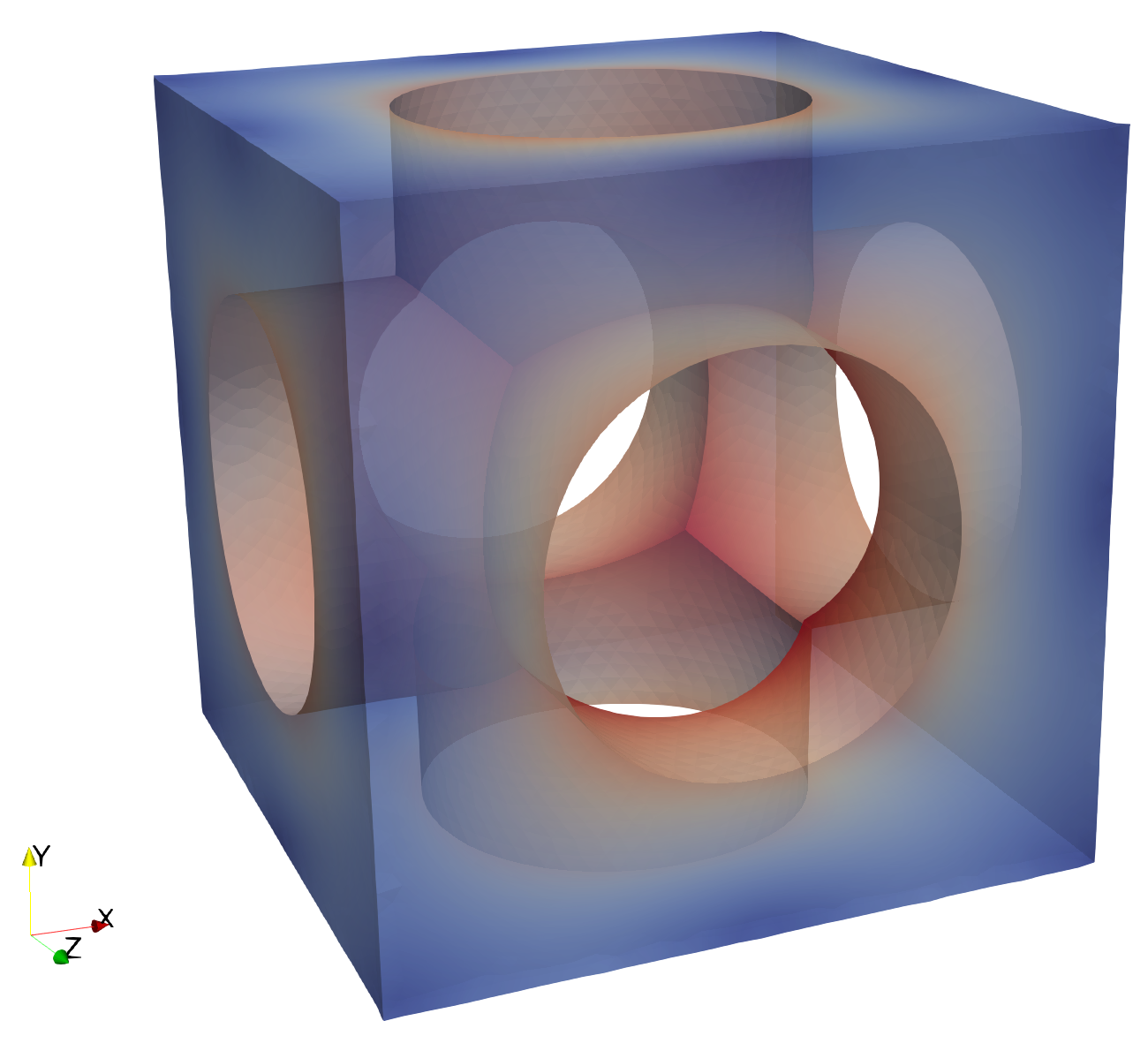}
	\centering
	\caption{Displacement (magnitude of $\vect u^{(1)}$) field of the RVE due to the pore pressure ($p^{(0)}$) in the absence of any macroscopic deformation ($\nabla_{\vect X} \vect u^{(0)}$). The higher displacement magnitude at the intersection of the three cylinders is notable.}
	\label{disp_pressure}
	\end{subfigure} \quad
	\begin{subfigure}{\iWidth cm}
	\includegraphics[width= \iWidth cm]{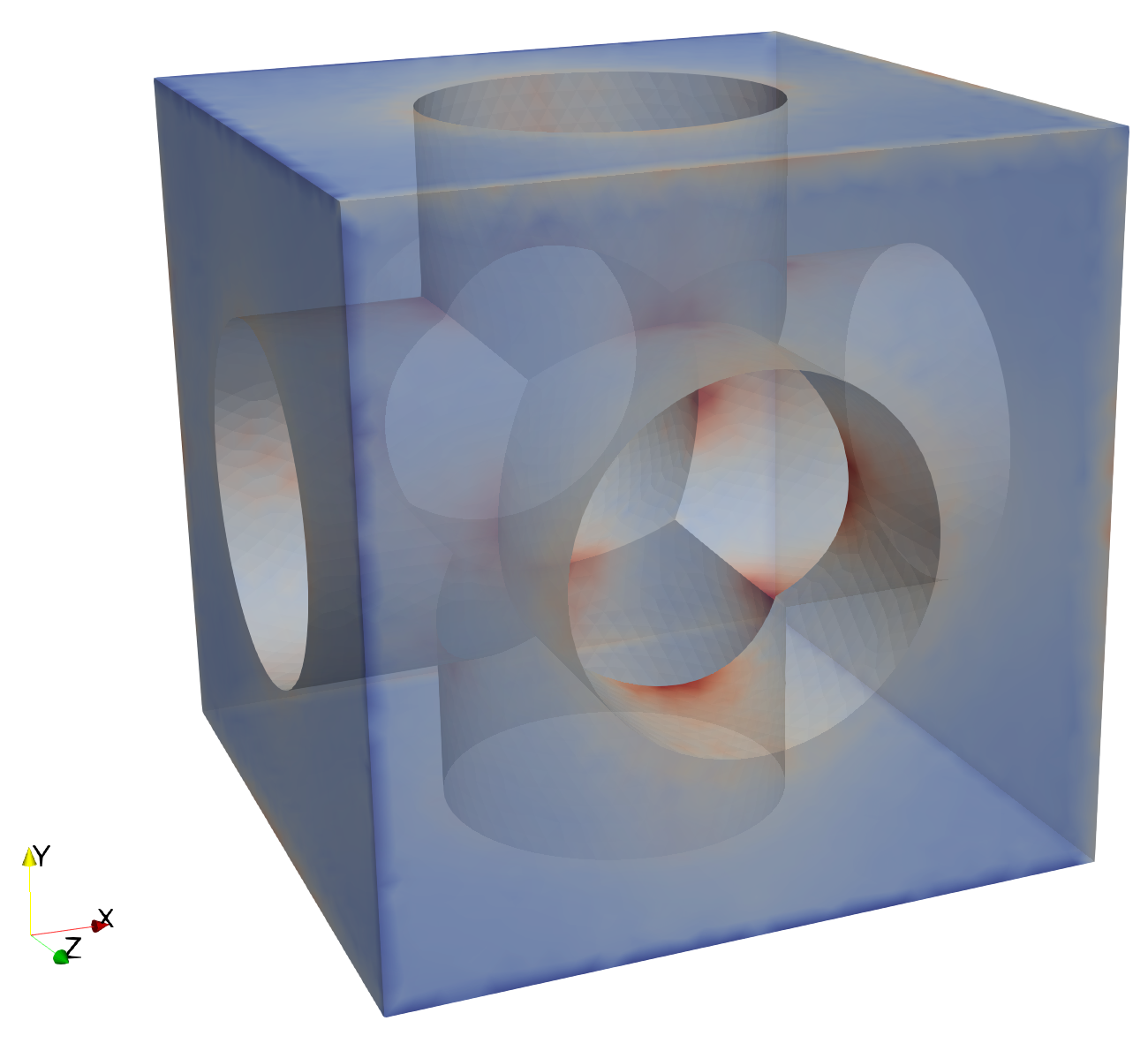}
	\centering
	\caption{The strain energy density field of the RVE corresponds to the same BCs as in Figure \ref{disp_pressure}. The energy concentration at the intersection of the fluid channels is of utmost importance, particularly, in the context of fracture/damage analysis.}
	\label{energy_pressure}
	\end{subfigure}
\caption{In this figure and Figures \ref{disp_energy_uni} and \ref{disp_energy_shear}, the displacement field is projected in the deformed Lagrangian configuration while the strain energy density field is projected in the reference Lagrangian configuration. The colour maps from blue to red show the magnitude variations from small to large values.}
\label{disp_energy_pressure}
\end{figure*}	

\begin{figure*}
\centering
	\begin{subfigure}{\iWidth cm}
	\includegraphics[width=  \iWidth cm ]{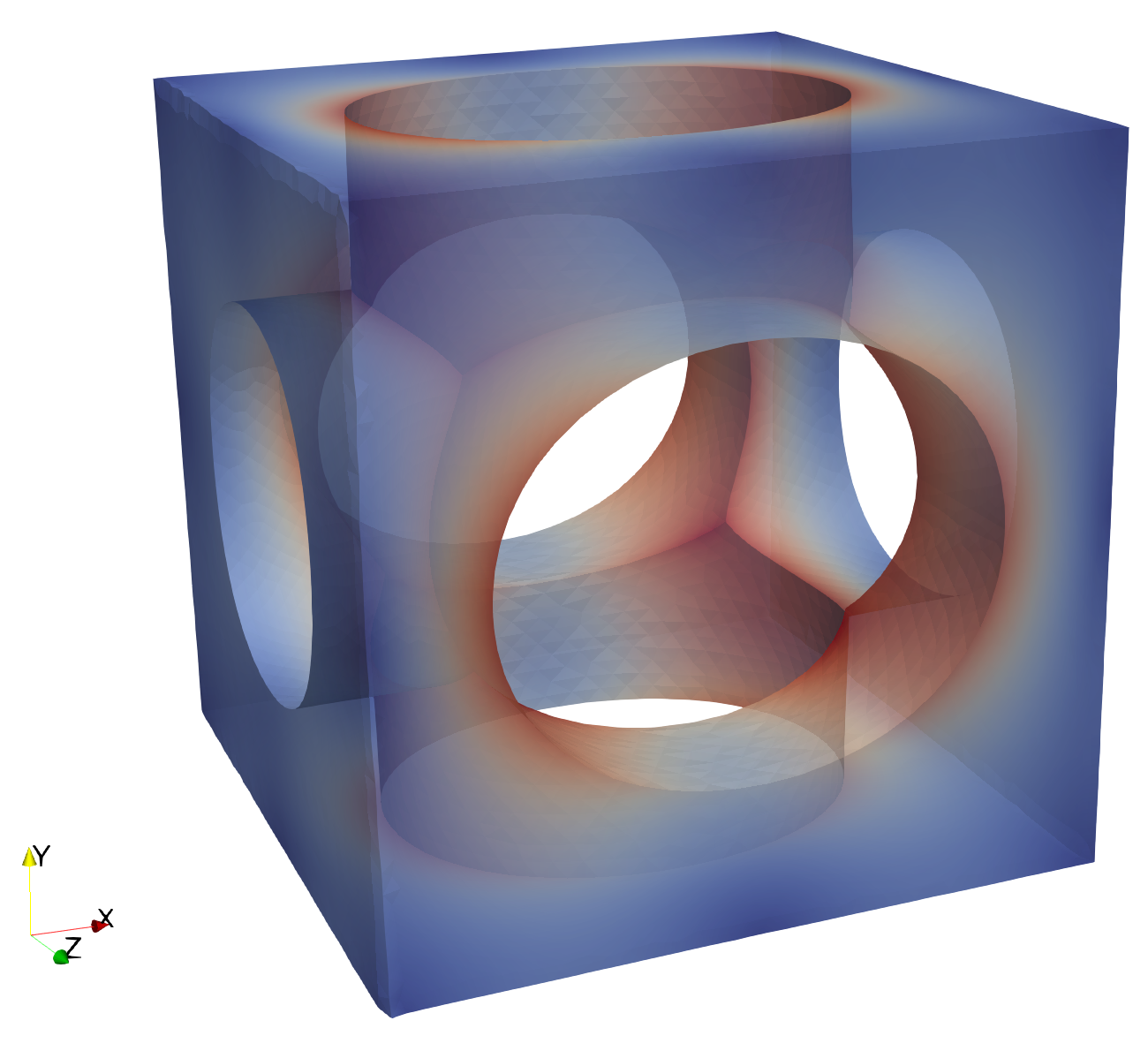}
	\centering
	\caption{The microscale magnitude field of $\vect u^{(1)}$ projected in the deformed Lagrangian configuration showing an elongation along the X-axis which takes place under the macroscopic uniaxial displacement gradient ($\nabla_X u^{(0)}_{11}$) in the absence of any other components.}
	\label{disp_uni}
	\end{subfigure} \quad
	\begin{subfigure}{\iWidth cm}
	\includegraphics[width= \iWidth cm]{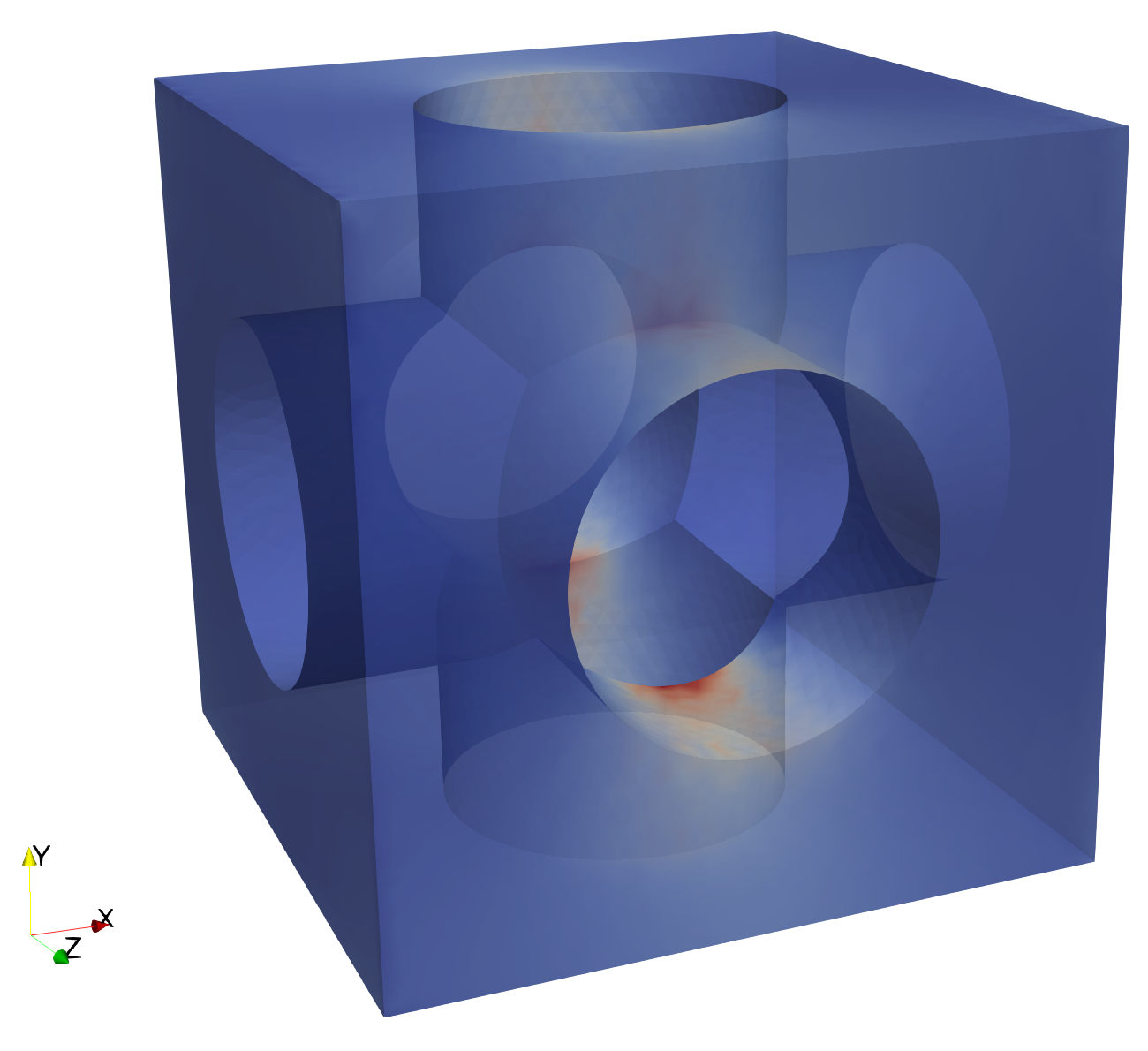}
	\centering
	\caption{The concentration of the strain energy density in the middle of the RVE length in the X-axis under $\nabla_X u^{(0)}_{11}$. \vspace{\LineS cm} \vspace{\LineS cm}\vspace{\LineS cm}\vspace{0.15 cm}}
	\label{energy_uni}
	\end{subfigure}
\caption{The microscale response fields of the poroelastic medium can be computed given the macroscale response of the matter.}
\label{disp_energy_uni}
\end{figure*}	

\begin{figure*}
\centering
	\begin{subfigure}{\iWidth cm}
	\includegraphics[width=  \iWidth cm ]{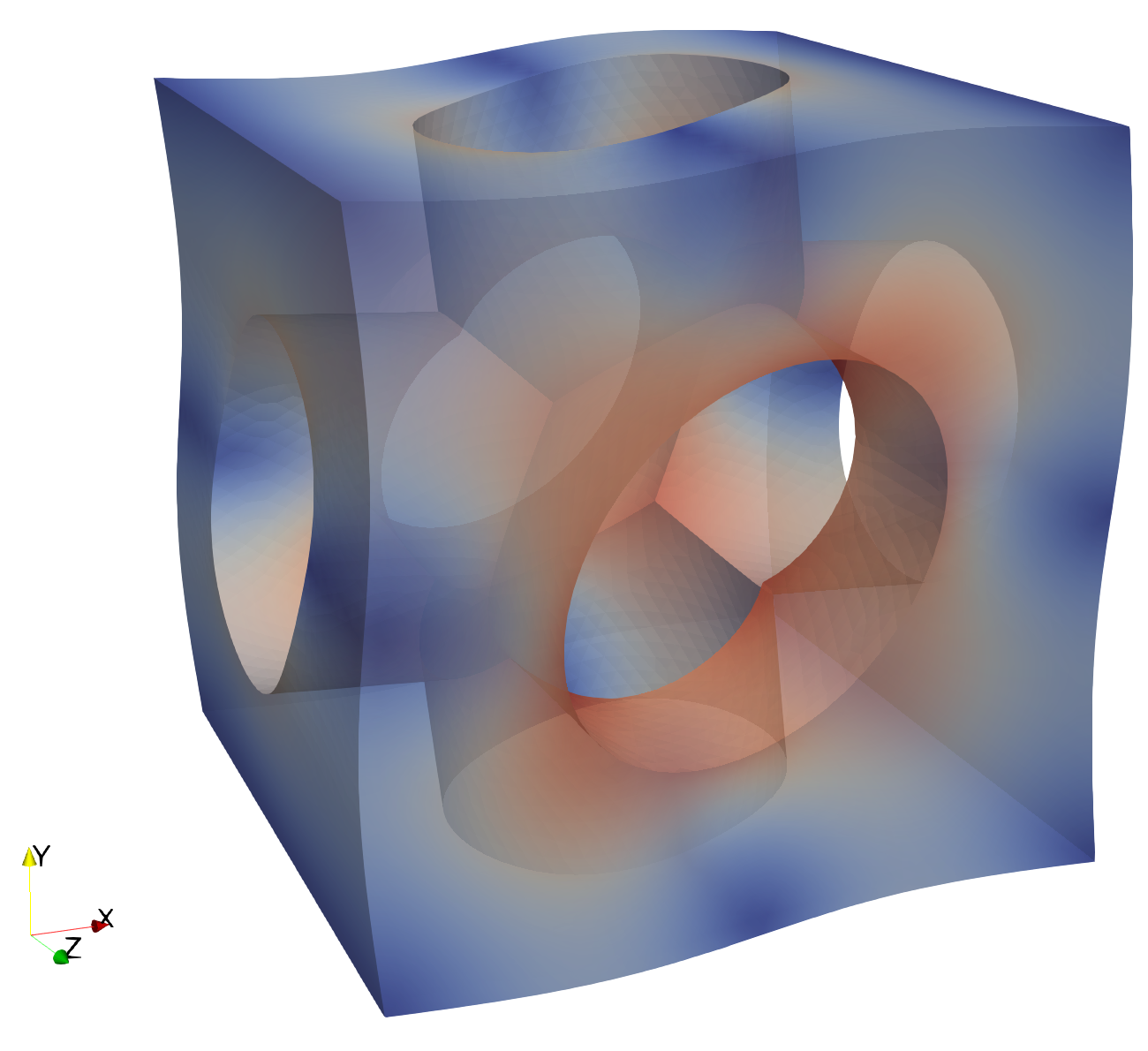}
	\centering
	\caption{The peculiar deformed configuration of the RVE under shear macroscale displacement gradient (more explicitly, $\nabla_X u^{(0)}_{12}$) is due to the assumed periodicity (equal inlet and outlet displacements) of the RVEs within one macroscale finite element.}
	\label{disp_shear}
	\end{subfigure} \quad
	\begin{subfigure}{\iWidth cm}
	\includegraphics[width= \iWidth cm]{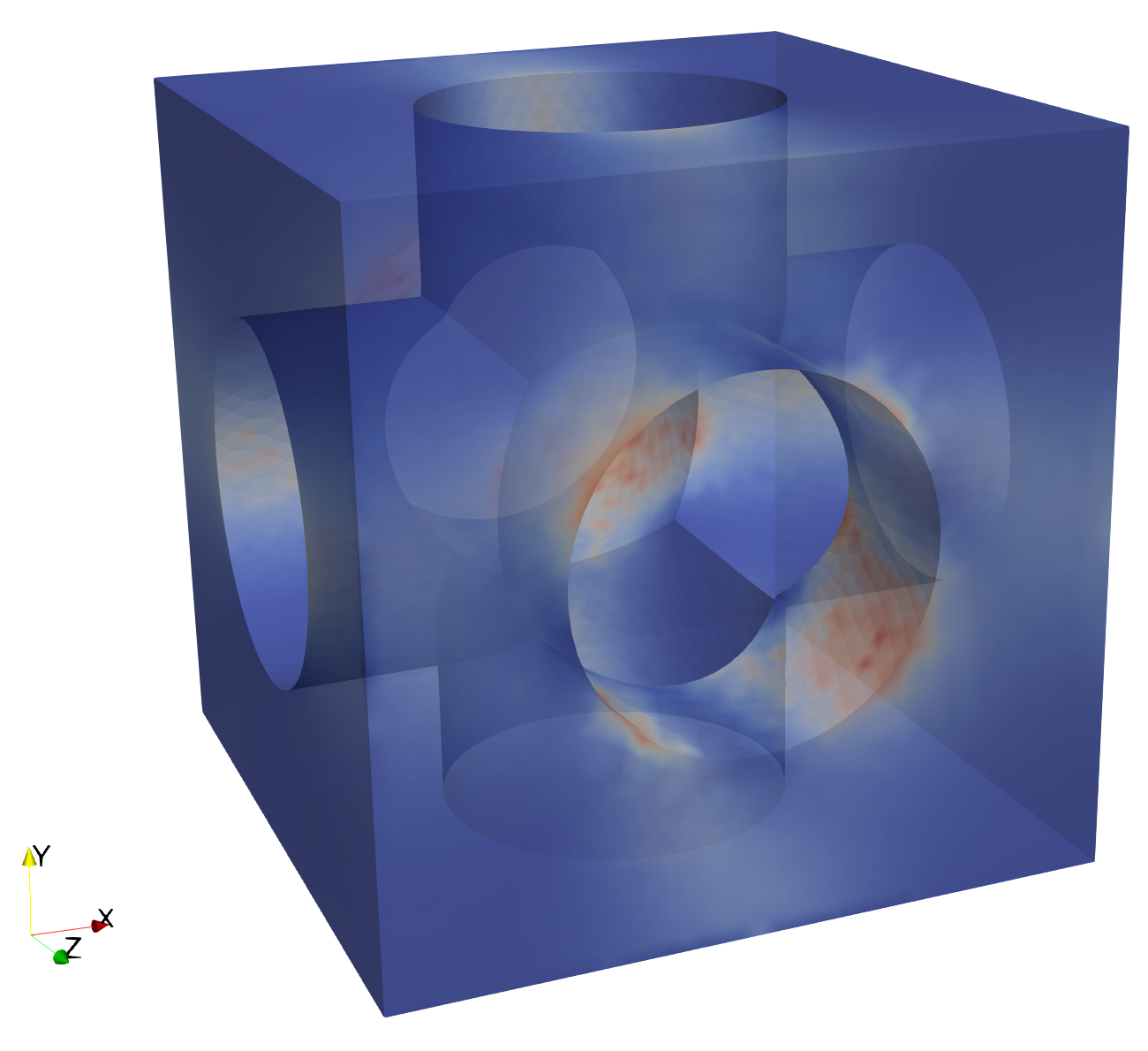}
	\centering
	\caption{The deformation illustrated in Figure \ref{disp_shear} results in the shown strain energy density field with a considerable concentration.\vspace{\LineS cm}\vspace{\LineS cm}\vspace{\LineS cm}}
	\label{energy_shear}
	\end{subfigure}
\caption{The microscale response under $\nabla_X u^{(0)}_{12}$.}
\label{disp_energy_shear}
\end{figure*}

\subsection{RVE numerical tests} \label{RVEnumtests}
Localisation, which is one of the important points provided by the present methodology, allows us to compute the mechanical response field of the microstructure under arbitrary macroscopic deformation and porepressure. This can be served to study phenomena such as local strain energy density (or deformation/stress) distribution and concentration (e.g. in the context of local fracture/damage mechanism). Here, we consider a unit cube as microstructure RVE. The solid phase is the cube from which three cylinders (the pores or fluid channels) with radius $r_f = 0.2$ are subtracted resulting in a porosity of $\phi = 0.29$ (for graphical representation of the RVE geometry see Figure \ref{energy_pressure}). The material model is the compressible neo-Hookean as in Equation \eqref{Psi_nonDim}. \textcolor{red}{Based on sensitivity analyses (minimising the dependency of the results on domain discretisation), the solid domain is discretised by 147043 tetrahedral elements with 25440 nodes. At this spatial resolution, solving a single solid problem (which takes 6.514 [s]) plus post-processing (which takes 14 [s]) of the response (projection of $\nabla_{\vect Y} \vect u^{(1)}$ in solid domain and calculation of its volumetric average) takes 20.51 [s] (wall time) using MacBook Pro with 2,4 GHz processor (with no parallel computing).}
 We compute the microscale response of the $\vect y$-periodic RVE under different macroscale conditions, namely, uniaxial ($\nabla_X u^{(0)}_{11}$), volumetric ($\nabla_X u^{(0)}_{11} = \nabla_X u^{(0)}_{22} = \nabla_X u^{(0)}_{33}$), and shear ($\nabla_X u^{(0)}_{12}$) macroscopic displacement gradients, as well as pore pressure ($p^{(0)}$). Since the latter conditions have strong interdependencies, we embrace the one-factor-at-a-time (OAT) sensitivity analysis method (varying only one macroscopic condition while the others are absent/zero) to study the role of each case. 	

\begin{figure*}
\centering
	\begin{subfigure}{\iWidth cm}
	\includegraphics[width=  \iWidth cm ]{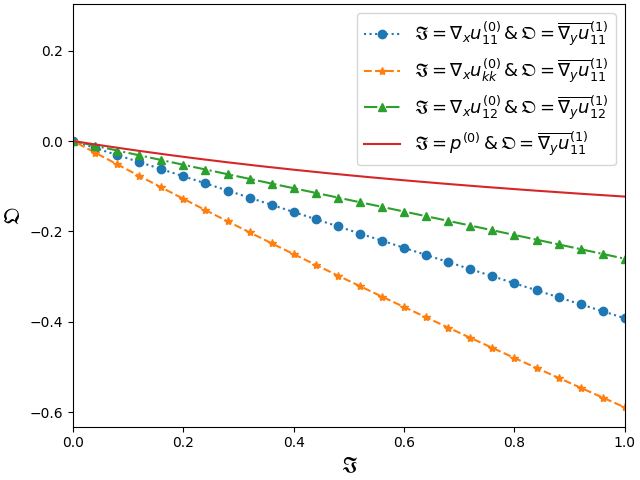}
	\centering
	\caption{"Corresponding elements" of average microscale displacement gradient tensor.}
	\label{dyu1_Av_1}
	\end{subfigure} \quad
	\begin{subfigure}{\iWidth cm}
	\includegraphics[width= \iWidth cm]{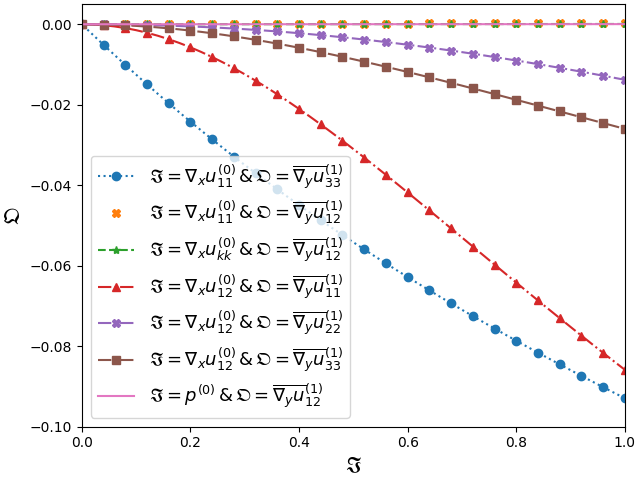}
	\centering
	\caption{"Non-corresponding elements" of average microscale displacement gradient tensor.}
	\label{dyu1_Av_2}
	\end{subfigure}
\caption{The profile of average microscale deformations under different components of the macroscale displacement gradient tensor $\nabla_{\vect X} \vect u^{(0)}$. The output ($\overline{\nabla_{\vect Y} \vect u}^{(1)}$) will be used as a part of the macroscale system of PDEs. Apart from the nonlinear profiles of different microscale response components, the considerable amount of diagonal components of RVE displacement gradient tensor ($\mathfrak{O} =  \nabla_Y u^{(1)}_{ii}$) under shear macroscale deformation ($\mathfrak{I} =  \nabla_X u^{(1)}_{12}$) show one of the major benefits of using the introduced methodology (robustness and accuracy). We highlight that the latter is neglected using Eulerian/linear poroelastic formulations (e.g. by enforcing zero $M_{ii12}$ in the calculation of effective stiffness tensor),  }
\label{dyu1_Av}
\end{figure*}

\begin{figure*}
\centering
	\begin{subfigure}{\iWidth cm}
	\includegraphics[width=  \iWidth cm ]{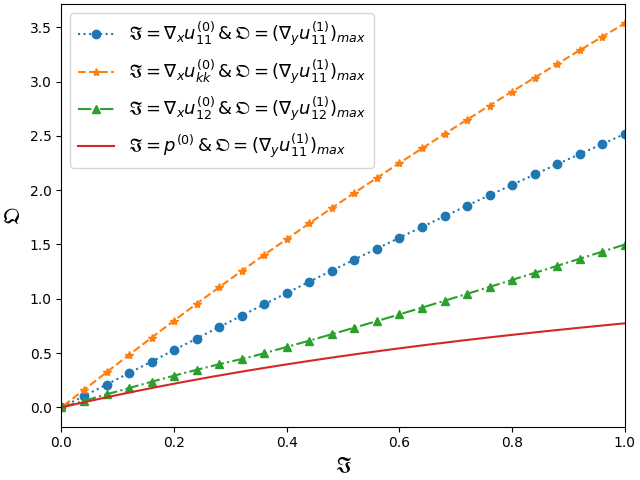}
	\centering
	\caption{"Corresponding elements" of maximum microscale displacement gradient tensor.}
	\label{dyu1_max_1}
	\end{subfigure} \quad
	\begin{subfigure}{\iWidth cm}
	\includegraphics[width= \iWidth cm]{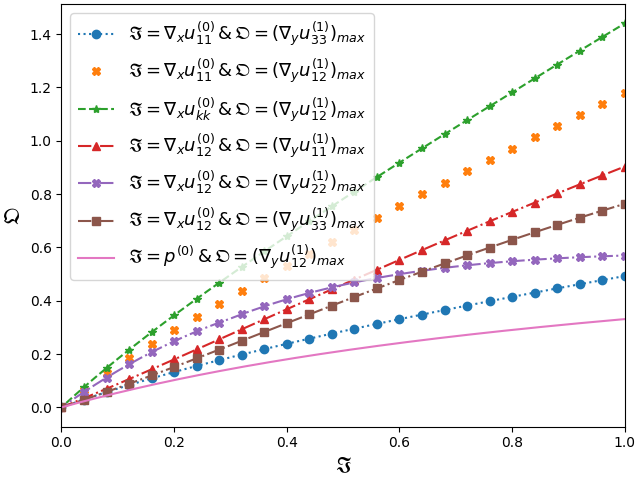}
	\centering
	\caption{"Non-corresponding elements" of maximum microscale displacement gradient tensor.}
	\label{dyu1_max_1}
	\end{subfigure}
\caption{Maximum microscale displacement gradient components showing the local deformation concentration. The significant difference between the average values shown in Figure \ref{dyu1_Av_1} and the maximum ones indicates the remarkable importance of computing the RVE's response field which is available in the present methodology. Furthermore, it is shown that, although the average values of several components in Figure \ref{dyu1_Av_2} are zero/negligible the maximum values are considerable with potential local effects.}
\label{dyu1_max}
\end{figure*}

\begin{figure*}
\centering
	\begin{subfigure}{\iWidth cm}
	\includegraphics[width=  \iWidth cm ]{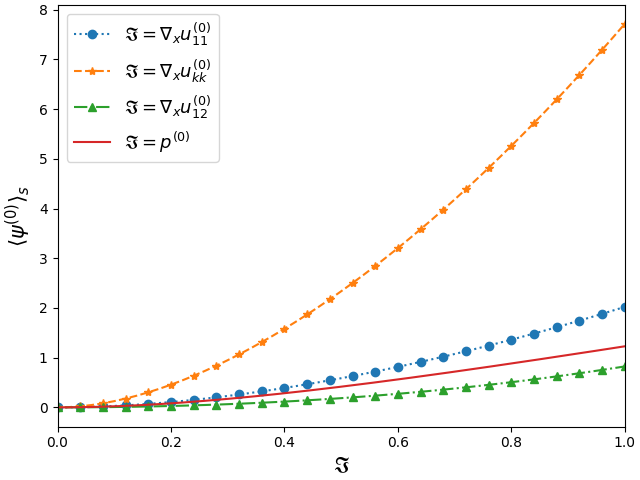}
	\centering
	\caption{Average strain energy density}
	\label{psi_Av}
	\end{subfigure} \quad
	\begin{subfigure}{\iWidth cm}
	\includegraphics[width= \iWidth cm]{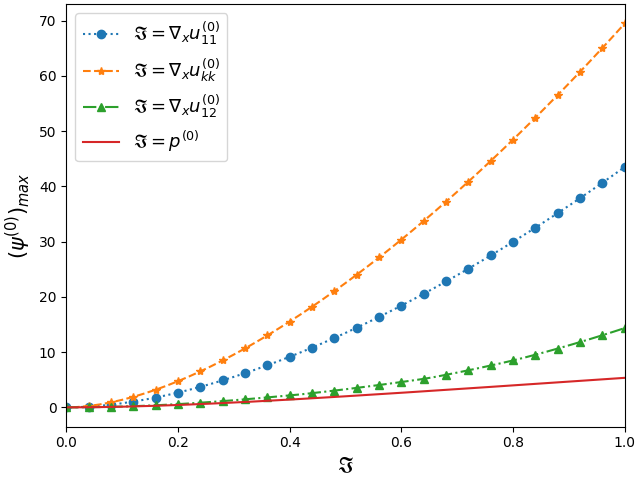}
	\centering
	\caption{Maximum strain energy density}
	\label{psi_max}
	\end{subfigure}
\caption{Strain energy density is a crucial parameter in damage/fracture analysis. The difference between average and maximum strain energy density (maximums are approximately 10 times greater than averages) shows that the local phenomena such as damage, fracture, etc. are probable although the macroscale analyses are far from the relevant thresholds.}
\label{psi}
\end{figure*}

As expected, the microscopic response is spatially heterogeneous with regions of high concentration. The colormaps of displacement magnitude and strain energy density shown in Figures \ref{disp_energy_pressure}, \ref{disp_energy_uni}, \ref{disp_energy_shear} visualises the local concentration zones of each field. In general, there are nine displacement gradient elements yet we only provide the profile of the ones with considerable effects on the overall response of the medium. The latter is divided into "Corresponding elements" and "Non-corresponding elements" which are shown in Figure \ref{dyu1_Av} (compare the indices of $\mathfrak{I}$ with $\mathfrak{O}$). Figures \ref{dyu1_Av} and \ref{dyu1_max} show, respectively, the variations of the average and maximum displacement gradient elements. Figure \ref{psi} provides the average and maximum strain energy density under different macroscopic conditions with further explanations provided in close captions. 

Hydraulic conductivity is a critical parameter in determining the hydraulic response of a poroelastic medium which varies in  cases under large porepressure/deformation. In the present methodology, this variation is taken into consideration using a transformation from the deformed configuration to the reference configuration. Considering the term $\frac{1}{J}$ in Equation \eqref{Vect_transform}, which is factorised and removed from mass conservation (in Equation \eqref{mass01} to simplify the mathematical procedure), and neglecting the effects of $\epsilon J^{(1)}$ in the expanded term $\frac{1}{J^{(0)} + \epsilon J^{(1)}}$, the full form of the transformed hydraulic conductivity takes the form
\begin{align}
\tens K_t = \frac{1}{J^{(0)}} \bar{\tens G}^{(0)}  \tens K  (\bar{\tens F}^{(0)})^{-T}. \label{transformed_hydro}
\end{align}


 \begin{figure*}
\centering
	\begin{subfigure}{\iWidth cm}
	\includegraphics[width= \iWidth cm]{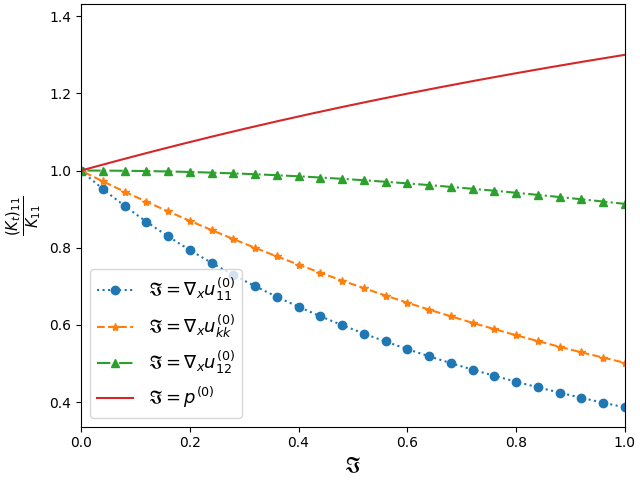}
	\centering
	\caption{Normalised transformed hydraulic conductivity ($(K_t)_{11}$ w.r.t the initial one).}
	\label{K11}
	\end{subfigure} \quad
	\begin{subfigure}{\iWidth cm}
	\includegraphics[width=  \iWidth cm ]{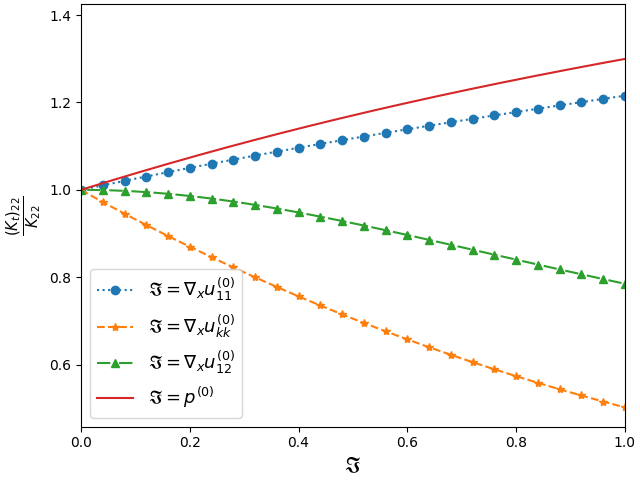}
	\centering
	\caption{Normalised transformed hydraulic conductivity ($(K_t)_{22}$ w.r.t the initial one).}
	\label{K22}
	\end{subfigure} 
\caption{Hydraulic conductivity is transformed using the leading order deformation gradient tensor composed of macroscale and macroscale displacement gradient tensors.}
\label{Hydraulic}
\end{figure*}
 
\begin{figure*}
\centering
	\begin{subfigure}{\iWidth cm}
	\includegraphics[width=  \iWidth cm ]{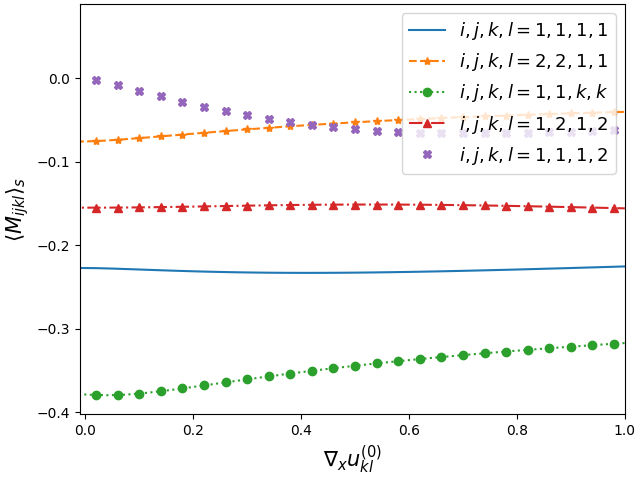}
	\centering
	\caption{The fourth order tensor $\langle \Tens M \rangle_s = \frac{\partial \langle\nabla_{\vect Y} \vect u^{(1)}\rangle_s}{\partial \nabla_{\vect X} \vect u^{(0)}}$ provides a part of FEM Jacobian matrix of the macroscale problem. In linear infinitesimal and remodelling-based finite strain poroelasticity (see e.g. \cite{Hdehghani, HDAZporo2021}, respectively), this tensor determines the poroelastic parameters including effective elasticity tensor and Biot coefficient. \vspace{\LineS cm}}
	\label{Mijkl_reference}
	\end{subfigure} \quad
	\begin{subfigure}{\iWidth cm}
	\includegraphics[width= \iWidth cm]{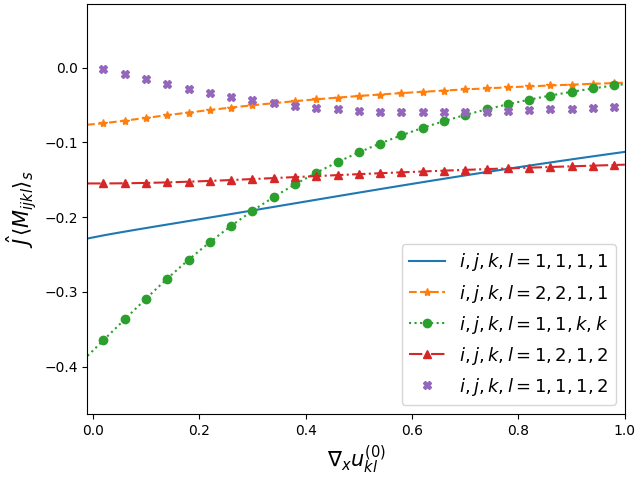}
	\centering
	\caption{The integral average in the deformed configuration is calculated by multiplying the RVE volumetric strain ($\hat J = \det(\nabla_{\vect Y} \vect u^{(1)} + \tens I) $) to the integral average of $\mathbb M$. The larger variations in this plot show that solving a poroelastic problem in reference configuration results in considerably smaller changes in the Jacobian matrix rendering the problem numerically more stable.\vspace{0.22 cm}}
	\label{Mijkl_deformed}
	\end{subfigure}
	\begin{subfigure}{\iWidth cm}
	\includegraphics[width=  \iWidth cm ]{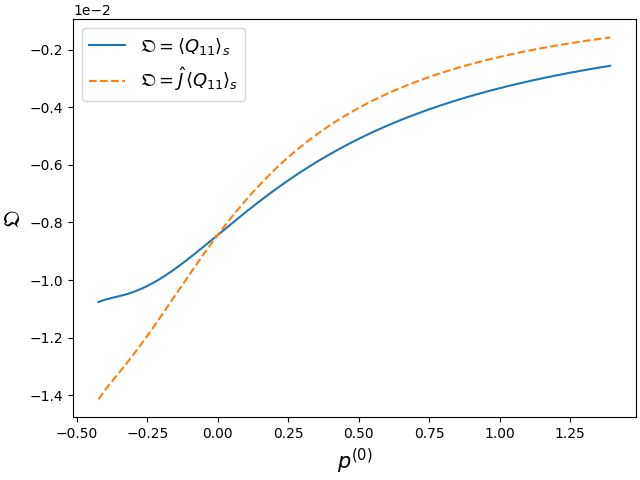}
	\centering
	\caption{The second order tensor $\tens Q = \frac{\partial \langle\nabla_{\vect Y} \vect u^{(1)}\rangle_s}{\partial p^{(0)}}$ also provides a part of FEM Jacobian matrix of the macroscale problem. In this plot, similar to Figures \ref{Mijkl_reference} and \ref{Mijkl_deformed}, it is shown that using ALE formulation the problem will be numerically more stable.}
	\label{Q_11}
	\end{subfigure} \quad
	\begin{subfigure}{\iWidth cm}
	\includegraphics[width= \iWidth cm]{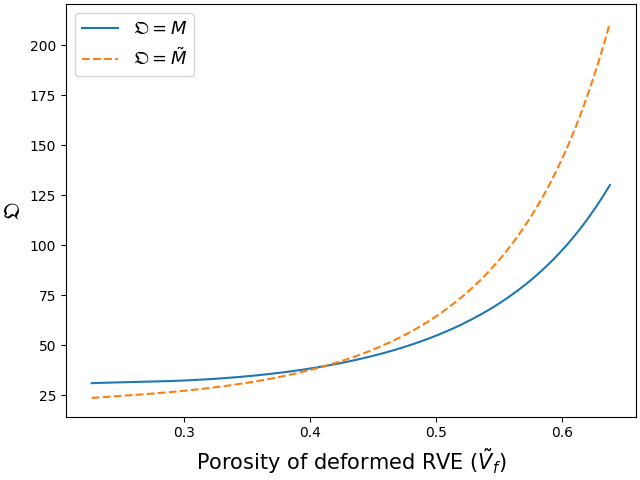}
	\centering
	\caption{Biot modulus $M =\frac{-1}{\Tr \langle \tens Q\rangle_s}$ (and transformed in Lagrangian deformed configuration $\tilde M =\frac{-1}{\hat J \Tr  \langle \tens Q\rangle_s}$) is one of the poroelastic parameters that could be compared with the results in \cite{Hdehghani}. \vspace{\LineS cm}\vspace{\LineS cm}\vspace{\LineS cm}}
	\label{Biot_Modulus}
	\end{subfigure}
\caption{Plots of poroelastic parameters $\langle \mathbb M \rangle_s$, $\langle \tens Q\rangle_s$, and $M$ showing that the problem with ALE formulation results in less variations in FEM Jacobian matrix.}
\label{Poroelastic}
\end{figure*}

 The variations of the main elements of the transformed hydraulic conductivity (normalised with respect to its initial value) due to the prescribed macroscale conditions and the resultant microscale effects are provided in Figure \ref{Hydraulic}. Finally, for the sake of comparison with linear poroelasticity, the profiles of the forth order tensor $\Tens M$, the second order tensor $\tens Q$, and Biot modulus (which are constants in linear poroelasticity) under $\nabla_{\vect X} \vect u^{(0)}$ and $p^{(0)}$ are provided in Figure \ref{Poroelastic}.
 
\subsection{Data-driven approach using ANNs} \label{sec_ANN}
In this section, we introduce ANNs as a surrogate for the solid cell problems that provide micro-macro scales link. We choose ANNs for this task because of the complexities enforced by the strong interrelation between the pore pressure and macroscale deformation gradient. This interrelation is due to, firstly, the pore pressure Neumann B.C. (introduced via Equation \eqref{SolidcellBC}) which depends on the leading order Piola transformation, secondly, the macroscale displacement gradient which results in a body force-type load and, thirdly, a neo-Hookean hyperelastic model which is nonlinear. Thus, a simple superposition of the effects is not valid.

\begin{figure*}
\centering
	\includegraphics[width=  9  cm ]{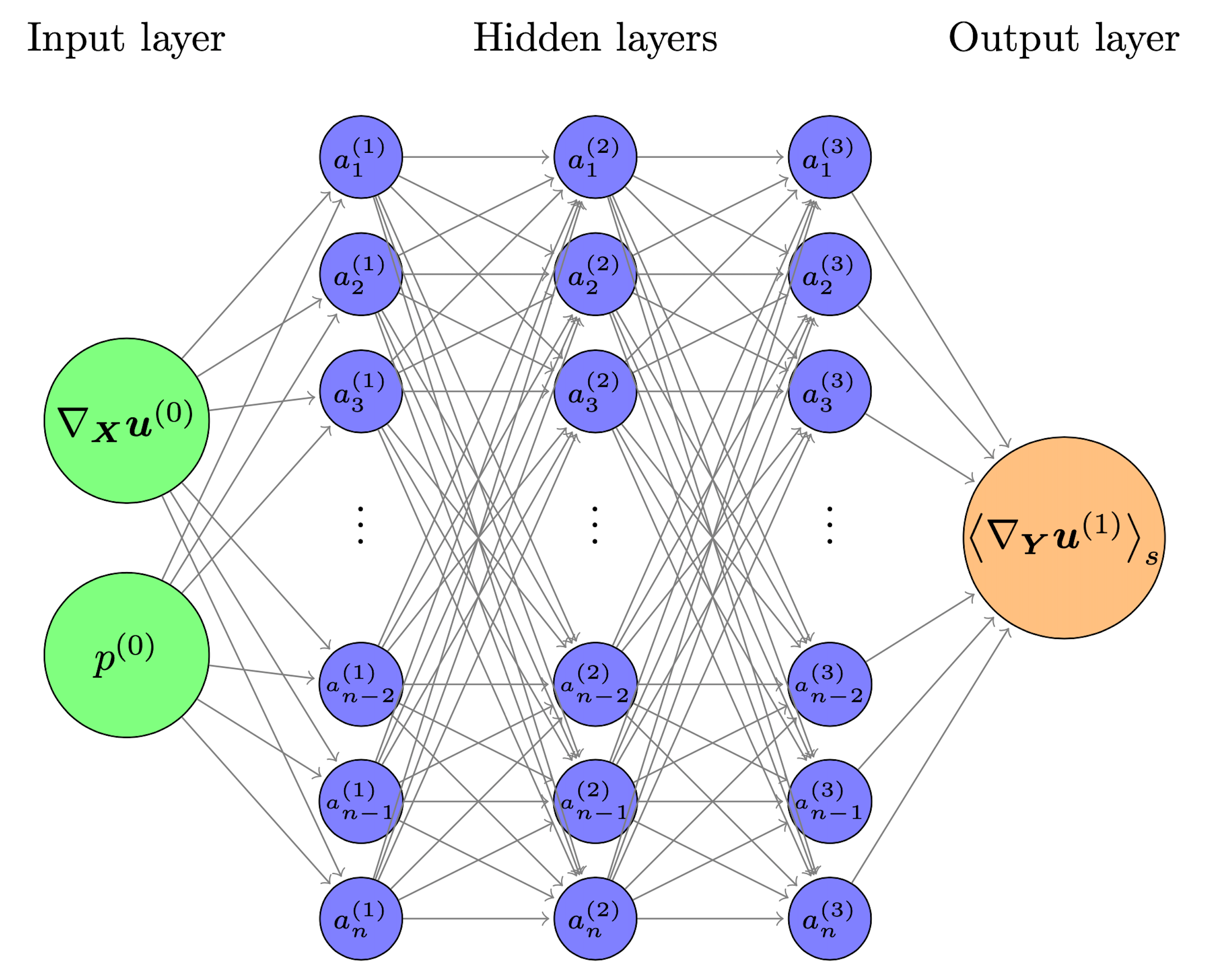}
	\centering
	\caption{Schematic representation of $\langle\nabla_{\vect Y} \vect u^{(1)}\rangle_s = \text{ANN}(\nabla_{\vect X} \vect u^{(0)},p^{(0)})$ as a surrogate model for microscale solid RVE problem. We highlight that this ANN is only valid for a specific geometry of RVE, material model, and material parameters.}
	\label{ANN_config}
\end{figure*}

\textcolor{red}{The RVE problems solved via DNS are too time-consuming to be included directly in the calculations of the provided concurrent multi-scale multi-physics approach. However, they are useful to provide dataset for training ANNs.} Basically, the aim is to replace a continuous problem represented by a system of PDEs (RVE problem to be solved via DNS \textcolor{red}{with a geometry described in Section \ref{RVEnumtests}}) with a surrogate model (here, ANNs) that provides the response faster than the original approach.
In general, the inputs of the ANN are the parameters affecting the microscopic response $\nabla_{\vect Y} \vect u^{(1)}$ including geometrical features, material parameters, and macroscale inputs ($\nabla_{\vect X} \vect u^{(0)}$ and $p^{(0)}$). However, since providing a dataset for the general case is computationally expensive we adopt a case-specific approach in which only the varying parameters are introduced as the inputs of ANNs (see Figure \ref{ANN_config} for a schematic graphical representation of the ANN). In this study, for verification/benchmark and comparison purposes, we adopt the relevant varying inputs of a confined consolidation test (e.g. Terzaghi's test), namely, $\frac{\partial u^{(0)}_2}{\partial X_2}$ and $p^{(0)}$ with the outputs being $\frac{\partial u^{(1)}_1}{\partial Y_1}$, $\frac{\partial u^{(1)}_2}{\partial Y_2}$, and $\frac{\partial u^{(1)}_3}{\partial Y_3} \approx \frac{\partial u^{(1)}_1}{\partial Y_1}$. More complex examples will be considered in future studies. \textcolor{red}{The suitable complexity of the model is chosen via ANN Hyperparameters tuning (so-called, Grid Search). In fact, we choose the smallest possible ANN architecture that delivers accurate results to 1- calculate the outputs efficiently and 2- to avoid overfitting due to unnecessarily large ANN capacity. The chosen ANN architecture has three hidden layers with 40 Neurones with Sigmoid activation functions in each layer. Furthermore, it is trained using Adam optimiser and MSE loss function.}

 Despite their efficiency, the accuracy/fidelity of deep learning approaches as the surrogate models are directly linked to the density (or sample rate) of the discrete sequence (training dataset) so that it reflects the features of the original function. This is provided by solving a sufficient number of cell problems described via Equation \eqref{Solidcell} and \eqref{SolidcellBC}. 
 The "sufficient" number of the problems to be solved (samples of training dataset) to provide the required density of the training dataset largely depends on the original problem. In case the latter shows a varying polynomial degree of nonlinearity (which is the case, here) the optimum density will be heterogeneous. Furthermore, covering the space of  the high number of input elements at a given interval requires taking all the possible combinations into account. This renders an extreme computational cost if blind sampling techniques (such as equidistant discretisation at the finest required steps) are adopted.
 
Here, we develop a simple real-time output density check to \textcolor{red}{adaptively refine the output density (adaptive sampling-step refinement). This ensures that the provided dataset reflects the features of the continuous problem at an optimum density so that the well-trained ANN will deliver the desired accuracy. }Although the latter is not the focus of this study we briefly explain it. 
Let us adopt a simple equidistantly sequenced sampling through incremental analysis of the RVE solid problem with one moving input element at a time (so-called one-factor-at-a-time OAT). With a default step/increment size of $\Delta_i^d \nabla_{\vect X} \vect u^{(0)}$ and $\Delta_i^p p^{(0)}$ we define the residual of sample $n$ as

\begin{align}
{\rm Res}_{max}^n = \frac{{\rm max} (\lvert(\nabla_{\vect Y} \vect u^{(1)})^n - (\nabla_{\vect Y} \vect u^{(1)})_L^n\rvert)}{\lvert\lvert \nabla_{\vect Y} \vect u^{(1)} \rvert\rvert_F}, \label{Res}
\intertext{ where $\lvert\lvert \tens \sbt \rvert\rvert_F = [\sum_{i,j} \sbt_{i,j}^2]^{1/2}$ indicates the Frobenius norm. 
The term $(\nabla_{\vect Y} \vect u^{(1)})_L^n$ is the result of a linear extrapolation which can be described by}
(\nabla_{\vect Y} \vect u^{(1)})_L^n = J_{n-1} \Delta {\rm I}_n + (\nabla_{\vect Y} \vect u^{(1)})_L^{n-1},
\intertext{where $\Delta I_n$ is the variation in the moving element of input array ($\Delta {\rm I}_n = {\rm I}_n - {\rm I}_{n-1}$) and $J_{n-1}$ is the tangent acquired based on the results of increments $n-2$ and $n-1$ using}
((\nabla_{\vect Y} \vect u^{(1)})_L^{n-1} - (\nabla_{\vect Y} \vect u^{(1)})_L^{n-2})/\Delta {\rm I}_{n-1}.
\end{align}
Finally, if the residual is more than a pre-defined tolerance the moving input element will adopt the value $${\rm I}_{n'} = \frac{{\rm I}_{n} + {\rm I}_{n-1}}{2}.$$

 \begin{figure}
\centering
	\includegraphics[width=  6  cm ]{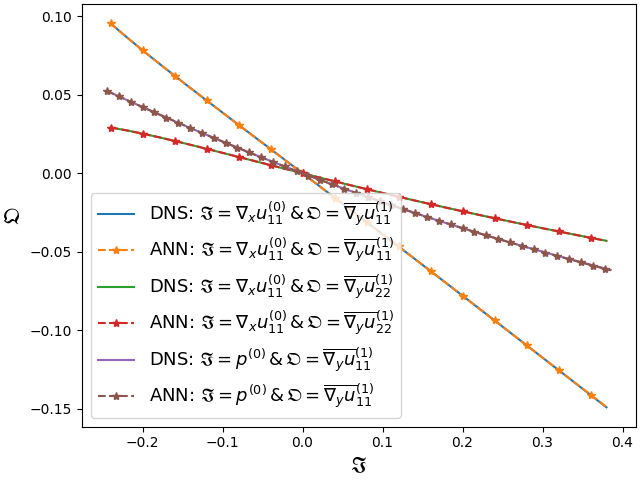}
	\centering
	\caption{ANN test}
	\label{ANN_test}
\end{figure} 

\textcolor{red}{A training dataset suitable for the numerical reconstruction of a consolidation test (introduced in the following section) with two input variables is to be provided. To this end, we choose relatively coarse initial  sampling steps (0.03 for $-0.4 \leq \frac{\partial u^{(0)}_2}{\partial X_2} \leq 0.4$ and 0.1 for  $-1 \leq p^{(0)} \leq 1$ (non-dimensional), which are adaptively refined by the described real-time output density check, wherever required. The number of solved solid problems (described by Equation \eqref{SolidRVEweak}) sums to a total of 1189, which, given the required time for each problem in Section \ref{RVEnumtests}, takes 406,44 minutes. The required time to train the ANN can vary, but it is usually around 10 minutes.} 

\textcolor{red}{
The results of the provided ANN is valid for many cases including any characteristic RVE size, solid material constant, and fluid viscosity (since they are non-dimensionalised). However, in cases with different parameters whose variations are not considered in this model (e.g. porosity in \cite{HDAZ2020}) a more comprehensive dataset and network architecture is required.}

\paragraph{\textcolor{red}{On the computational speed-up}}
\textcolor{red}{Solving a solid problem to achieve the required results for homogenised model $\langle\nabla_{\vect Y} \vect u^{(1)}_t\rangle_s$ using DNS (as described in Section \ref{RVEnumtests}) takes around 20.5 [s] while, employing the provided ANN, it takes around 0.0002 [s]. For comparison purposes, in the simple consolidation problem described in Section \ref{numericalEX}, the ANN is called 2250 times per each time increment ($\approx 0.45$ [s] per increment). Solving one increment of this problem without ANN (using DNS) could take nearly 13 hours in our MacBook Pro with 2,4 GHz Processors. It is noteworthy that, depending on the problem, achieving the steady-state solution of a poroelastic problem requires fine time discretisation (e.g. 200 time-increments).}

 \subsection{Macroscale problem} \label{macroweak}
Equations \eqref{Equi_macro}, \eqref{effectiveP}, \eqref{AverageP0}, \eqref{Eqmass2}, and \eqref{Darcy_ALE} construct a system of PDEs describing the macroscopic/effective problem.
The weak formulation of the latter reads
\begin{align} 
\nonumber
{0} =& + \int_{\Omega_h} \tens P_E {:} \nabla_{\vect X}\delta \vect u^{(0)} dV - \int_{\partial \Omega_h} \vect t \cdot \delta \vect u^{(0)} {\rm d}S \\ \nonumber
 & + \int_{\Omega_h} \langle \bar{\tens G}^{(0)}  \vect w \rangle_f  \cdot \nabla_{\vect X}\delta p^{(0)} dV \\ \nonumber
 & - \int_{\partial \Omega_h} \langle \bar{\tens G}^{(0)}   \vect w \rangle_f \cdot \vect N_h \delta p^{(0)} {\rm d}S  \\ \nonumber
 & + \int_{\Omega_h} \langle \bar{\tens G}^{(0)T} {:}\nabla_{\vect Y} \dot{ \vect u}^{(1)} \rangle_s \delta p^{(0)} dV \\
 &- \int_{\Omega_h} \left\langle \bar{\tens G}^{(0)T}  {:} \nabla_{\vect X}   \dot{\vect u}^{(0)} \right\rangle_f \delta p^{(0)} {\rm d}V,
\end{align}
where $\delta \vect u^{(0)}$ and $\delta p^{(0)}$ are the required test functions.
Since the problem is history and time-dependent we need to perform an incremental analysis which reads 
\begin{align}\nonumber
{0} =& + \int_{\Omega_h} \Delta \tens P_E {:} \nabla_{\vect X}\delta \vect u^{(0)} dV - \int_{\partial \Omega_h}  \Delta \vect t \cdot \delta \vect u^{(0)} {\rm d}S \\ \nonumber
 & + \int_{\Omega_h} \Delta \langle \bar{\tens G}^{(0)}   \vect w \rangle_f \cdot \nabla_{\vect X}\delta p^{(0)} dV \\ \nonumber
 & - \int_{\partial \Omega_h} \Delta \langle \bar{\tens G}^{(0)}   \vect w \rangle_f \cdot \vect N_h \delta p^{(0)} {\rm d}S  \\ \nonumber
 & + \int_{\Omega_h} \Delta \langle \bar{\tens G}^{(0)T} {:} \nabla_{\vect Y} \dot{ \vect u}^{(1)} \rangle_s \delta p^{(0)} dV \\ 
 & - \int_{\Omega_h}  \Delta \langle \bar{\tens G}^{(0)T} \rangle_f  {:} \nabla_{\vect X}   \dot{\vect u}^{(0)} \delta p^{(0)} {\rm d}V \label{weak_form}
\end{align}
where $\vect N_h$ is the normal vector to the surface of interest at macroscopic level in reference Lagrangian configuration. In fact, the problem reduces to finding $p^{(0)}_{t+\Delta t}$ and $\vect u^{(0)}_{t+\Delta t}$ such that Equation \eqref{weak_form} is fulfilled.

At this stage, it is necessary to calculate the increments of each field (stress, fluid velocity etc.) based on the increments of $\vect u^{(0)}$ and $p^{(0)}$, where $\vect u^{(0)}_{t+\Delta t} = \vect u^{(0)}_t + \Delta \vect u^{(0)}$ and $p^{(0)}_{t+\Delta t} = p^{(0)}_t + \Delta p^{(0)}$

\begin{align}
\nonumber
\Delta \tens P_E &= V_s \left(\frac{\partial \Psi^{(0)}}{\partial \bar{\tens F}^{(0)}_{t+\Delta t}} - \frac{\partial \Psi^{(0)}}{\partial \bar{\tens F}^{(0)}_t  }\right) - \\ 
  & \hspace{0.6cm}V_f \left(p^{(0)}_{t+\Delta t} \bar{\tens G}^{(0)T}_{t+\Delta t} - p^{(0)}_t \bar{\tens G}^{(0)T}_t\right).
\end{align}
Furthermore, $\langle\nabla_{\vect Y} \vect u^{(1)}_t\rangle_s$ could be calculated at the intermediate configuration ( $\langle\nabla_{\vect Y} \vect u^{(1)}_t\rangle_s = \text{ANN}(\nabla_{\vect X} \vect u^{(0)}_t, p^{(0)}_t)$) employing the trained ANN as described in Section \ref{sec_ANN}. As the ANN input-output tangent is required for the macroscopic iterative solver (Newton-Raphson) we consider the following transformation for a small increment of microscopic deformation
\begin{align}
\Delta \langle\nabla_{\vect Y} \vect u^{(1)}\rangle_s = \Tens M {:} \Delta \nabla_{\vect X} \vect u^{(0)} + \tens Q \Delta p^{(0)}, \label{linearisation}
\end{align}
where the fourth rank tensor $\Tens M$ and the second rank one $\tens Q$ could be estimated numerically at the end of the last increment using
\begin{align}
\Tens M = \frac{\text{ANN}(\nabla_{\vect X} \vect u^{(0)}_{t} + \delta \nabla_{\vect X} \vect u^{(0)},p^{(0)}_{t}) - {\langle\nabla_{\vect Y} \vect u^{(1)}_{t}\rangle_s}}{(\nabla_{\vect X} \vect u^{(0)}_{t} + \delta \nabla_{\vect X} \vect u^{(0)}) - \nabla_{\vect X} \vect u^{(0)}_{t}} \label{eqM1} \\
\tens Q = \frac{\text{ANN}(\nabla_{\vect X} \vect u^{(0)}_{t},p^{(0)}_{t}+ \delta p^{(0)}) - {\langle{\nabla_{\vect Y} \vect u}^{(1)}_{t}\rangle_s}}{(p^{(0)}_{t} + \delta p^{(0)}) - p^{(0)}_{t}}, \label{eqQ1}
\end{align}
where $ \delta \sbt$ (except for the test functions) in Equations \eqref{eqM1} and \eqref{eqQ1} indicate a very small variation in $\sbt$ so that taking the partial derivatives in a numerical way. 
Finally, the term $\Delta \langle \bar{\tens G}^{(0)}   \vect w \rangle_f$ could be calculated via

\begin{align}
\nonumber
\Delta \langle \bar{\tens G}^{(0)}   \vect w \rangle_f &=  - \bar{\tens G}^{(0)}_{t+\Delta t}  \tens K (\bar{\tens F}^{(0)}_{t+\Delta t})^{-T}\nabla_{\vect X}p^{(0)}_{t+\Delta t} \\
 & \hspace{1.2cm}+ \bar{\tens G}^{(0)}_t \tens K  (\bar{\tens F}^{(0)}_t)^{-T}\nabla_{\vect X}p^{(0)}_t
\end{align}

\section{Numerical example} \label{numericalEX}
We perform a confined compression (consolidation) test to, firstly, verify the implementation of the methodology and, secondly, to demonstrate to what extent employing the present model will improve the accuracy of the results when dealing with poroelastic/porohyperelastic problems under finite strain. Throughout this section, for the sake of abbreviation, we call the present method and the linear poroelastic equations the ALE and linear cases, respectively.  We solve the problem in a non-dimensional form and, at the end of this study, we dimensionalise the variables using the representative/unit values corresponding to soil mechanics and brain tissue applications.

\begin{figure}
\centering
\includegraphics[width=  7 cm ]{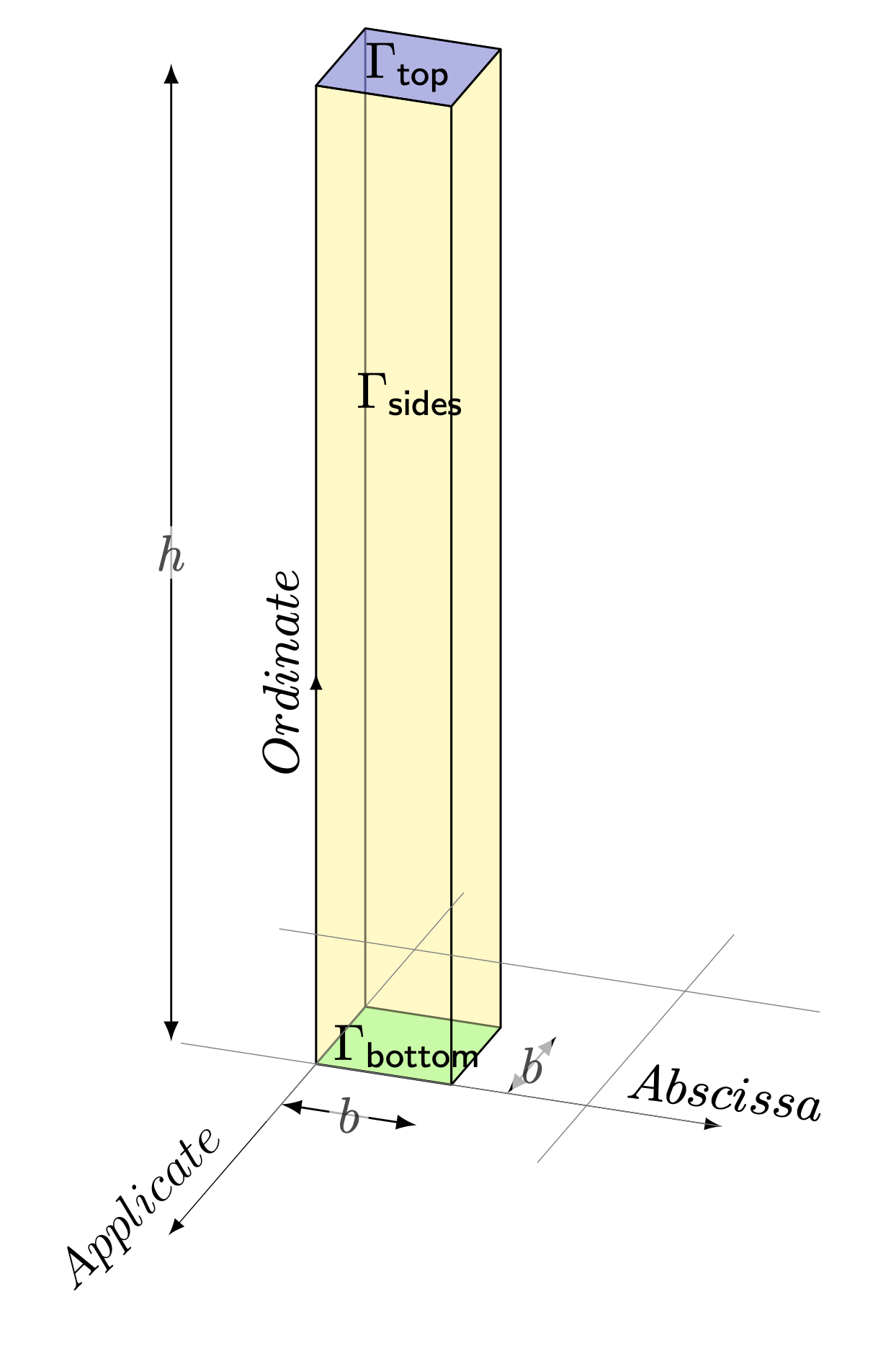}
\caption{Schematic representation of confined compression test with $h=7.5$ and $b = 0.1$. We use abscissa, ordinate, and applicate (corresponding indices, respectively, are 1, 2, and 3) to avoid confusion of coordinate axes with macroscale and microscale spatial variables.}
\label{Sketch_T}
\end{figure}
Let us assume that the column of poroelastic/poro-hyperelastic material shown in Figure \ref{Sketch_T} is under mechanical pressure $\vect P = P \vect N $ on $\Gamma_\mathsf{top}$ (the external load) where the fluid drainage is allowed. The column is impermeable on all other surfaces (namely,  $\Gamma_\mathsf{bottom}$ and  $\Gamma_\mathsf{sides}$) with zero displacement B.C. on $\Gamma_\mathsf{bottom}$ and zero displacement in $abscissa$ and $applicate$ coordinate axes but free displacement in $ordinate$ axis.

 \begin{figure*}
\centering
	\begin{subfigure}{\iWidth cm}
	\includegraphics[width=  \iWidth cm ]{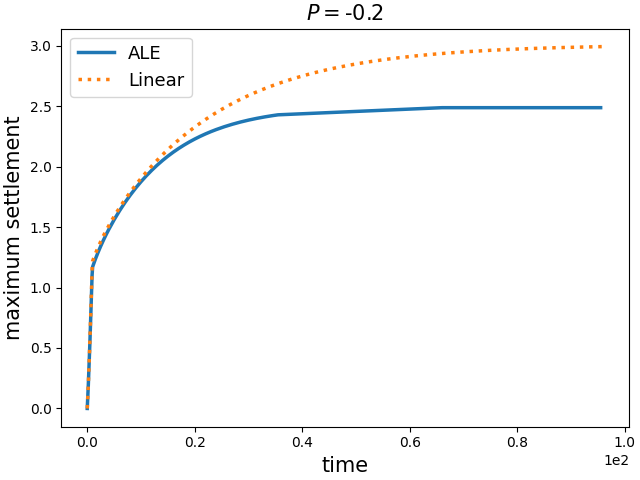}
	\centering
	\caption{As expected, the ALE case with neo-Hookean strain energy density function under finite strain exhibits stiffening behaviour resulting in the smaller final settlement. The settlement is around 30\% of the initial length of the column, thus transformation between deformed and undeformed configurations becomes of utmost importance.}
	\label{T_SettlementMax02}
	\end{subfigure} \quad
	\begin{subfigure}{\iWidth cm}
	\includegraphics[width= \iWidth cm]{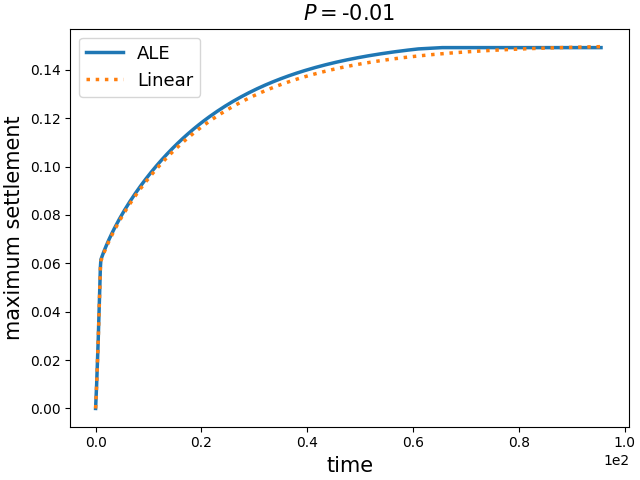}
	\centering
	\caption{The response of the ALE problem under small deformation (close to infinitesimal strain (around 2\%)) is very close to the linear poroelastic case, which is a source of verification. Although the final settlements are almost equal the ALE reaches the final settlement slightly earlier.\vspace{\LineS cm}\vspace{0.3 cm}}
	\label{T_SettlementMax001}
	\end{subfigure}
\caption{Maximum settlement takes place on $\Gamma_{top}$.}
\label{T_SettlementMax}
\end{figure*}

\textcolor{red}{We solve the mentioned problems using the present method with the neo-Hookean material model. We also solve the same problem using the linear poroelastic approach} with the effective elasticity tensor $\mathbb{\tilde C}$, Biot modulus $M$, Biot coefficient $\alpha$, and hydraulic conductivity $\tens K$ identified using the same RVE properties as the ALE model (for more details on multi-scale linear poroelasticity see e.g. \cite{Hdehghani1}). We expect the mechanical responses under smaller deformations to be close while they diverge at higher deformations. The dimensionless fluid dynamic viscosity $\mu^f = 1e{-3}$ is adopted resulting in initial hydraulic conductivity $(K_{i})_{jj} = 1.943 \quad j = 1, 2, 3$. \textcolor{red}{We discretise the homogenised domain with 106 elements (chosen through sensitivity analysis) in the longitudinal direction and one element in the other direction (with 3 nodal degrees of freedom) since the microscopic response $\langle\nabla_{\vect Y} \vect u^{(1)}\rangle_s$ is 3D although the homogenised response $\nabla_{\vect X} \vect u^{(0)}$ is 1D (see Figure \ref{dyu1_Av_2}).}

\begin{figure*}
\centering
	\begin{subfigure}{\iWidth cm}
	\includegraphics[width=  \iWidth cm ]{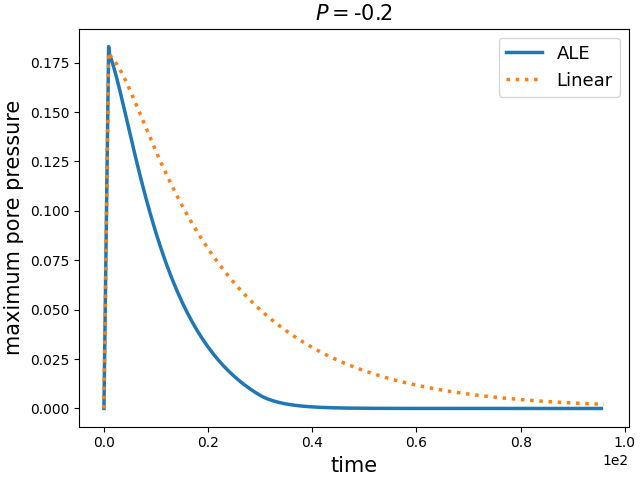}
	\centering
	\caption{The pore pressure in the ALE case vanishes faster than the linear one indicating a faster transition from the transient state to the steady-state, which could also be concluded from Figure \ref{T_SettlementMax02} where the ALE steady-state maximum settlement is reached quicker.}
	\label{T_PorepressureMax02}
	\end{subfigure} \quad
	\begin{subfigure}{\iWidth cm}
	\includegraphics[width= \iWidth cm]{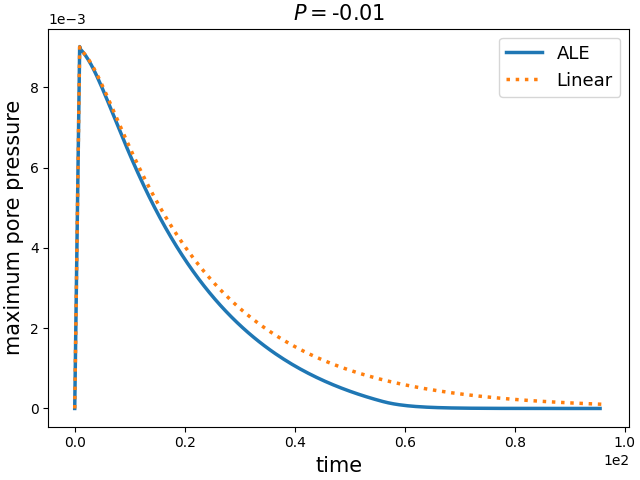}
	\centering
	\caption{The profile of pore pressure vs. time of the ALE and linear cases are closer under smaller deformations.\vspace{\LineS cm}\vspace{\LineS cm}\vspace{\LineS cm}\vspace{\LineS cm}}
	\label{T_PorepressureMax001}
	\end{subfigure}
\caption{Maximum pore pressure takes place on $\Gamma_{bottom}$ where it is the farthest place to the free drainage surface ($\Gamma_{top}$).}
\label{T_PorepressureMax}
\end{figure*}

We define settlement as the negative displacement in ordinate direction ($- u^{(0)}_2$). In order to ensure that the linearisation of the ANN (Equation \eqref{linearisation}) is valid/accurate, we apply the external load in 10 increments at the beginning of the consolidation process with very small changes in time (one-tenth of the usual time increments) which results in a sheer (but incremental) increase in settlement and porepressure (see Figures \ref{T_SettlementMax} and \ref{T_PorepressureMax}).

We highlight that the difference between the final/steady-state deformation of the ALE and the linear case reflects the deviation of the neo-Hookean hyperelastic material model from the linear elastic case while the Piola transformation seems to have a strong role in determining the quality of reaching there (with respect to time). Since at small deformations the Piola transformation tensor $\bar{\tens G}^{(0)}$ approaches the identity tensor and the neo-Hookean constitutive equation is approximately the same as the linear elastic one, the solid mechanical response of the ALE and linear cases are expected to be very close which is shown in Figure \ref{T_SettlementMax001}. The solid deformation is represented via maximum settlement on $\Gamma_\mathsf{top}$.  On the other hand, the ALE case in the test with $P = -0.2$ reaches a smaller settlement in a shorter time reflecting the effects of strain stiffening of the neo-Hookean material model under compression and the effects of Piola transformation.

\begin{figure*}
\centering
	\begin{subfigure}{\iWidth cm}
	\includegraphics[width=  \iWidth cm ]{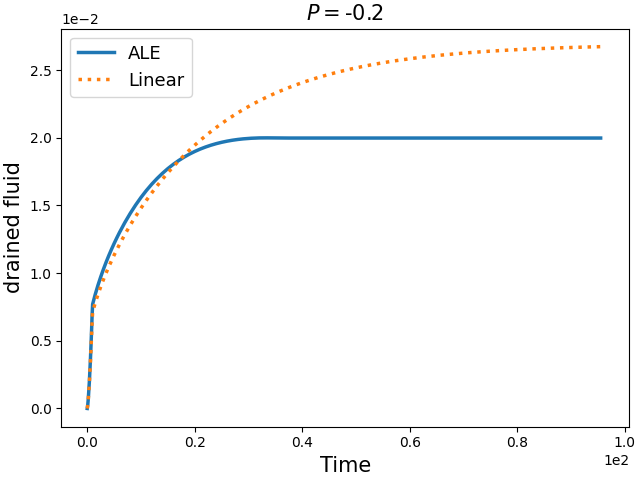}
	\centering
	\caption{Fluid drainage when the medium undergoes finite strain. In the ALE case the drainage takes place faster and the final drained fluid volume is considerably less than in the linear case.}
	\label{T_DrainedFluid02}
	\end{subfigure} \quad
	\begin{subfigure}{\iWidth cm}
	\includegraphics[width= \iWidth cm]{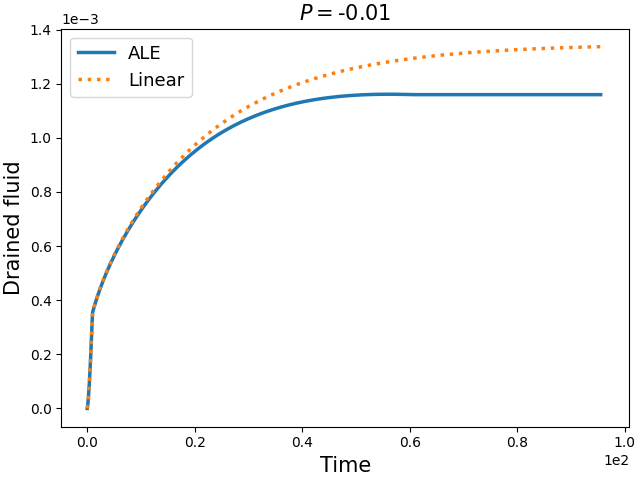}
	\centering
	\caption{The fluid drainage profile of ALE and linear cases under smaller deformation is closer compared to the Figure \ref{T_DrainedFluid02}.\vspace{\LineS cm}}
	\label{T_DrainedFluid001}
	\end{subfigure}
\caption{The fluid drainage is allowed only on $\Gamma_{top}$ where the mechanical BC ($P$) is applied. We calculate the drained fluid volume of ALE and linear cases, respectively, using the integrals $\int_{\Omega_h} \frac{1}{J^{(0)}} \nabla_{\vect X} \cdot (\bar{\tens G}^{(0)}  \vect w)  dV$ and $\int_{\Omega_h} \nabla_{\vect X} \cdot  \vect w_L  dV$, where $\vect w_L$ is the relative fluid velocity of the linear case. }
\label{T_DrainedFluid}
\end{figure*}

According to Equations \eqref{Eqmass2} and \eqref{Darcy_ALE}, the effect of Piola transformation and the leading order deformation gradient tensor $\bar{\tens F}^{(0)}$ on fluid flow and pore pressure profiles are of higher orders. Consequently, the deviation of pore pressure and drained fluid volume of ALE case from the linear one is distinguishable at even smaller settlements (at time 1.7e5 where the settlement is around 0.1) as shown in Figures \ref{T_PorepressureMax001} and \ref{T_DrainedFluid001}. This strong dependency is more evident in the profile of maximum pore pressure under the larger B.C. ($P = -0.2$)  shown in Figure \ref{T_PorepressureMax02} where the pore pressure vanishes considerably faster than the linear case.

\begin{figure*}
\centering
	\begin{subfigure}{\iWidth cm}
	\includegraphics[width=  \iWidth cm ]{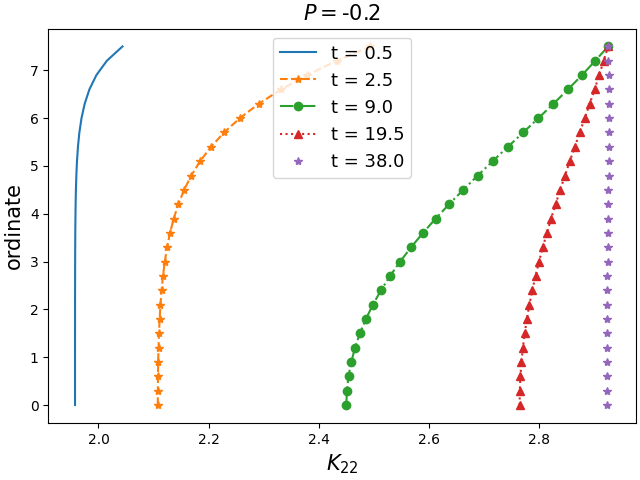}
	\centering
	\caption{The element $K_{22}$ increases during the consolidation process due to the transformation (dominated by $\nabla_X u^{(0)}_{22}$) in the reference configuration.\vspace{\LineS cm}}
	\label{y_K22}
	\end{subfigure} \quad
	\begin{subfigure}{\iWidth cm}
	\includegraphics[width= \iWidth cm]{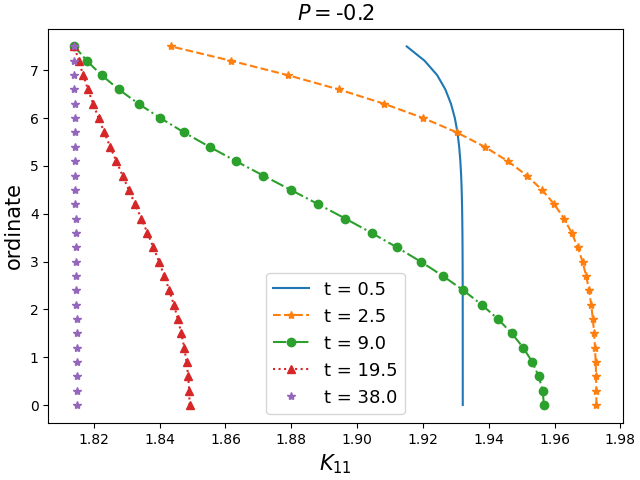}
	\centering
	\caption{The element $K_{11}$ increases at the bottom at shorter times (when the effects of pore pressure are dominant) while it decreases approaching the top of the column and at longer times.}
	\label{y_K11}
	\end{subfigure}
\caption{The profile of the main elements of hydraulic conductivity.}
\label{hydraulic_conductivity}
\end{figure*}

Concluding from Figures \ref{T_SettlementMax}, \ref{T_PorepressureMax}, and \ref{T_DrainedFluid}, although the nonlinear material law has a considerable effect in determining the final solid deformation, the effects of the microscale and macroscale solid deformation (which constitute $\bar{\tens G}^{(0)}$ and $\bar{\tens F}^{(0)}$) on the fluid part is even more critical. This is also reflected in the spatial profile of hydraulic conductivity at different times, shown in Figure \ref{hydraulic_conductivity}, where the transformed hydraulic conductivity varies according to the microscale and macroscale deformations with the initial value being $(K_{i})_{jj} = 1.943 \quad j = 1, 2, 3$.

 \begin{figure*}
\centering
	\begin{subfigure}{\iWidth cm}
	\includegraphics[width=  \iWidth cm ]{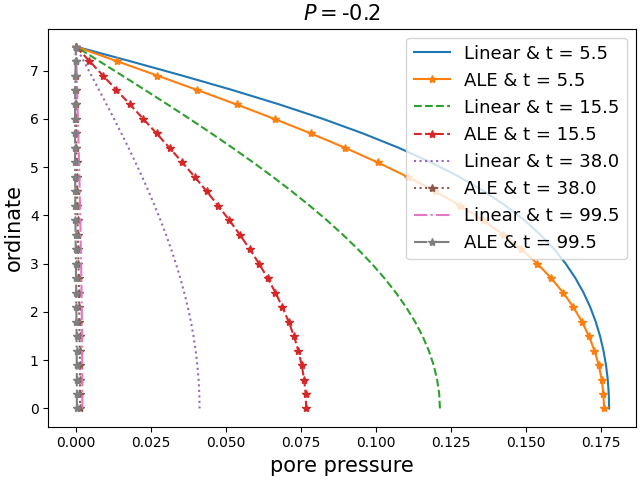}
	\centering
	\caption{Although shortly after the application of the Neumann BC the pore pressure spatial profile of ALE and linear cases are relatively close to each other they diverge at longer times, mainly due to the faster drainage in the ALE case.}
	\label{}
	\end{subfigure} \quad
	\begin{subfigure}{\iWidth cm}
	\includegraphics[width= \iWidth cm]{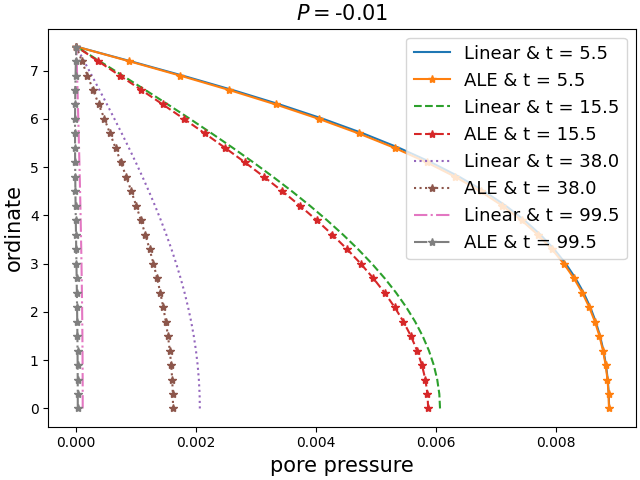}
	\centering
	\caption{The spatial pore pressure profiles of ALE and linear cases are closer when the applied Neumann BC is small.\vspace{\LineS cm}\vspace{\LineS cm}\vspace{\LineS cm}\vspace{\LineS cm}}
	\label{y_Porepressure}
	\end{subfigure}
\caption{The variations of pore pressure along the ordinate direction (Y-axis) at three different times, one shortly after the application of the load, one close to the steady-state, and one in between showing the transition from application of the BC until the final condition. We highlight that at the steady-state of this problem the pore pressure approaches zero everywhere in the space.}
\label{spatial_porepressure}
\end{figure*}

 \begin{figure*}
\centering
	\begin{subfigure}{\iWidth cm}
	\includegraphics[width=  \iWidth cm ]{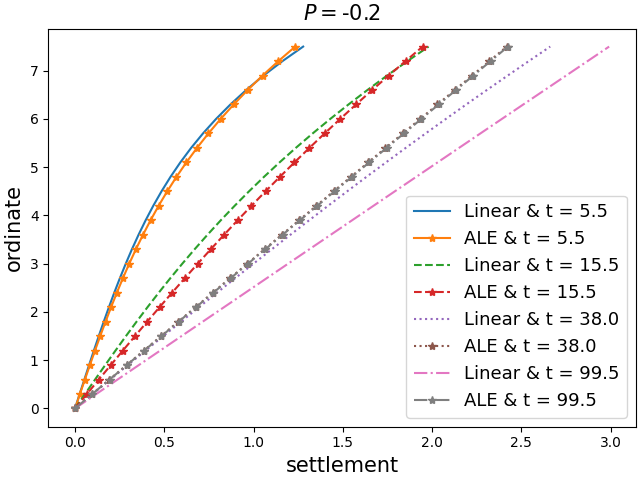}
	\centering
	\caption{Agreeing with Figure \ref{T_SettlementMax02}, the settlement profile of ALE diverges from the linear case. It reaches steady-state considerably sooner (at $t = 38$) than the linear case. Furthermore, the final settlement of the ALE case is around 20\% smaller.}
	\label{y_Settlement02}
	\end{subfigure} \quad
	\begin{subfigure}{\iWidth cm}
	\includegraphics[width= \iWidth cm]{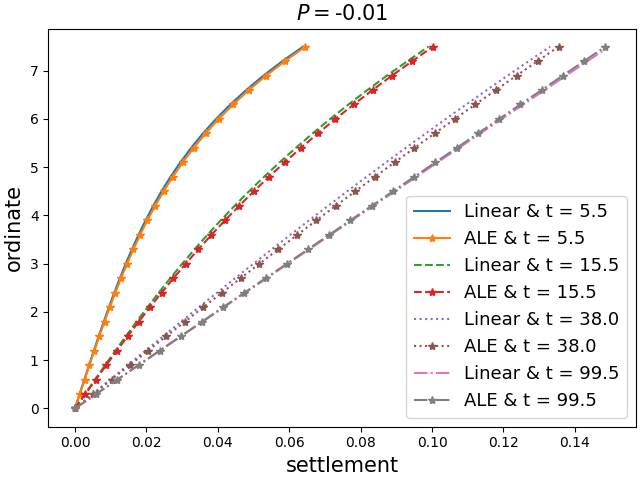}
	\centering
	\caption{Although the final settlements of both cases under small deformation are, more or less, the same, the settlement of the ALE case, during the transient state, is slightly more which is due to faster fluid drainage.\vspace{\LineS cm}}
	\label{y_Settlement001}
	\end{subfigure}
\caption{Settlement profile in ordinate direction. The final settlement profile at steady-state can indicate how the solid deformation is affected by the nonlinear material law while the time at which the settlement reaches the final state is more related to the fluid flow which is corrected using the leading order deformation gradient tensor.}
\label{y_Settlement}
\end{figure*}

\begin{figure}
\centering
	\includegraphics[width=  \iWidth cm]{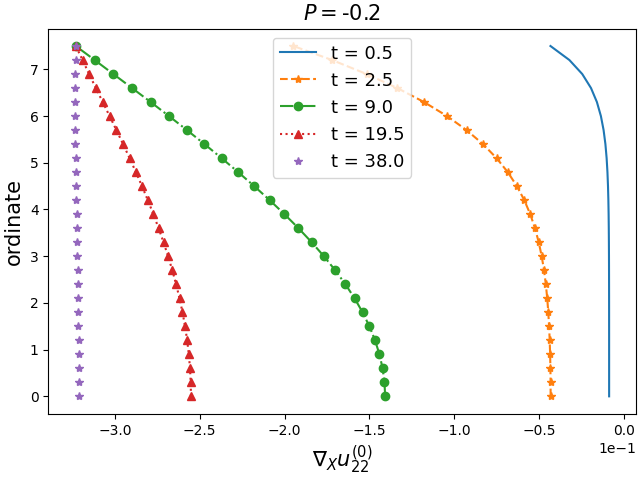}
	\caption{Since the compression test is confined, the main element of the macroscale displacement gradient is $\nabla_X u^{(0)}_{22}$. We highlight that the other elements are negligible compared to this one.}
	\label{y_dxu022}
\end{figure}

\begin{figure*}
\centering
	\begin{subfigure}{\iWidth cm}
	\includegraphics[width=  \iWidth cm ]{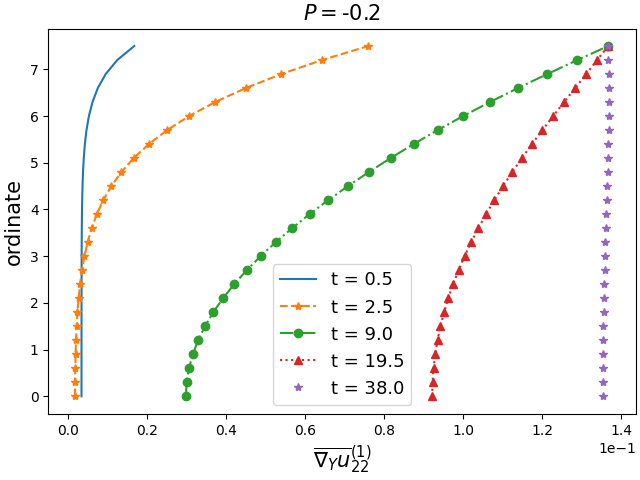}
	\centering
	\caption{$\overline{\nabla_Y u}^{(1)}_{22}$ is the main component of the average microscopic displacement gradient tensor. Although according to Figure \ref{dyu1_Av_1}, the positive pore pressure should result in a negative average microscopic displacement gradient, it is opposed by the effects of the negative values of $\nabla_X u^{(0)}_{22}$ enforcing positive average microscopic displacement gradient.}
	\label{y_dyu122}
	\end{subfigure} \quad
	\begin{subfigure}{\iWidth cm}
	\includegraphics[width= \iWidth cm]{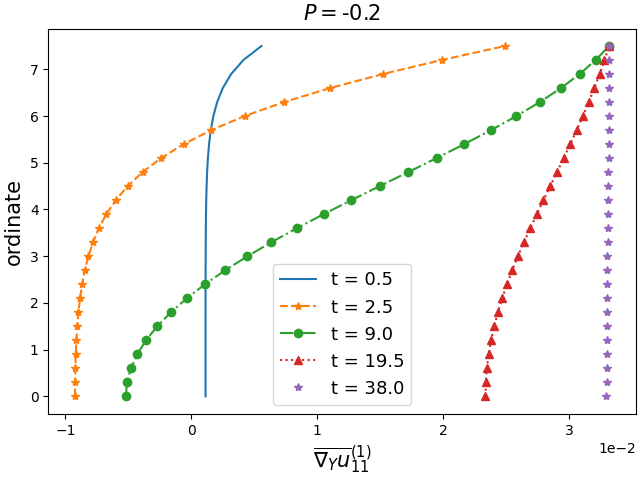}
	\centering
	\caption{$\overline{\nabla_Y u}^{(1)}_{11}$ is also a considerable element in determining the leading order deformation gradient tensor. The negative values at shorter times are due to the pore pressure while the positive values stem from the negative $\nabla_X u^{(0)}_{22}$ which is in agreement with Figures \ref{dyu1_Av_1} and \ref{dyu1_Av_2}.\vspace{\LineS cm}\vspace{\LineS cm}}
	\label{y_dyu111}
	\end{subfigure}
\caption{The considerable elements of $\overline{\nabla_{\vect Y} \vect u}^{(1)}$ ($\overline{\nabla_Y u}^{(1)}_{33}$ is similar to $\overline{\nabla_Y u}^{(1)}_{11}$) showing the combined effects of macroscale displacement gradient and pore pressure on microscale displacement gradient tensor.}
\label{y_dyu1}
\end{figure*}

The spatial profiles of the pore pressure, settlement, macroscopic and microscopic displacement gradients at different times are shown, respectively, in Figures \ref{spatial_porepressure}, \ref{y_Settlement}, \ref{y_dxu022}, and \ref{y_dyu111}. 
Comparing Figures \ref{spatial_porepressure}, \ref{y_Settlement001}, and \ref{y_dxu022} indicates that the pore pressure results in an increase in hydraulic conductivity in all directions (since it is a hydrostatic pressure) while the macroscopic displacement gradient $\nabla_X u^{(0)}_{22}$ (which is negative) increases $K_{22}$ but decreases $K_{11}$.

\begin{figure*}
\centering
	\begin{subfigure}{\iWidth cm}
	\includegraphics[width=  \iWidth cm ]{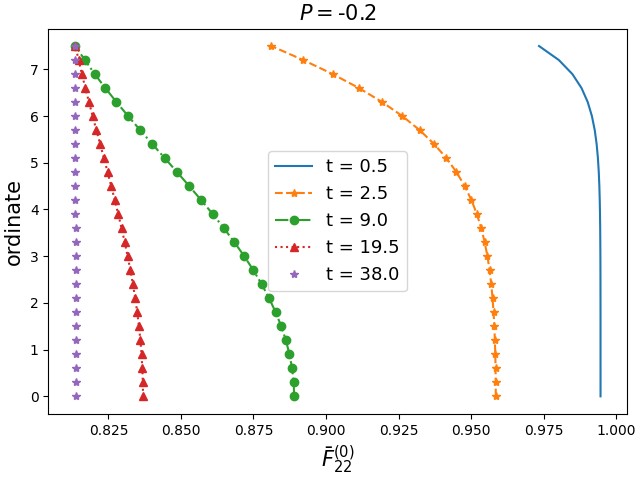}
	\centering
	\caption{Although, as shown in Figure \ref{y_dxu022}, the maximum of $\nabla_X u^{(0)}_{22}$ is around $-0.32$ the value of $\bar F^{(0)}_{22} = \nabla_X u^{(0)}_{22} + \overline{\nabla_Y u}^{(1)}_{22} + 1$ is moderated due to the opposite signs of $\nabla_X u^{(0)}_{22}$ and $\overline{\nabla_Y u}^{(1)}_{22}$.}
	\label{y_F022}
	\end{subfigure} \quad
	\begin{subfigure}{\iWidth cm}
	\includegraphics[width= \iWidth cm]{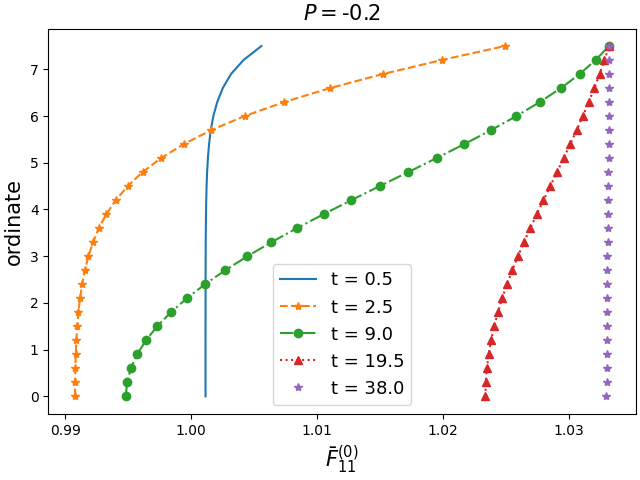}
	\centering
	\caption{The element $\bar F^{(0)}_{11}$ is only due to $\overline{\nabla_Y u}^{(1)}_{11}$ since the element $\nabla_X u^{(0)}_{11}$ is negligible.\vspace{\LineS cm}\vspace{\LineS cm}\vspace{\LineS cm}}
	\label{y_F011}
	\end{subfigure}
\caption{The existence of $\bar F^{(0)}_{11}$ and the difference between $\bar F^{(0)}_{22}$ and $\nabla_X u^{(0)}_{22} + 1$ shows the utmost importance of employing the provided real-time multi-scale methodology when dealing with finite strain poroelastic problems.}
\label{y_F0}
\end{figure*}

The macroscale deformation and pore pressure result in the microscale deformation (represented by $\overline{\nabla_{\vect Y} \vect u}^{(1)}$) whose spatial profiles at different times are provided in Figure \ref{y_dyu1}. The macroscale and microscale displacement gradients constitute the leading order deformation gradient which can be studied via Figure \ref {y_F0}. 

\begin{figure*}
\centering
	\begin{subfigure}{\iWidth cm}
	\includegraphics[width=  \iWidth cm ]{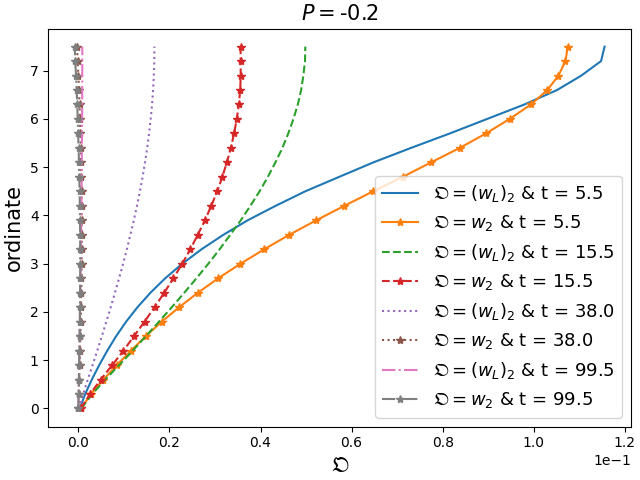}
	\centering
	\caption{ALE and linear relative fluid flow, similar to other parameters, diverge considerably.\vspace{\LineS cm}\vspace{\LineS cm}\vspace{\LineS cm}\vspace{\LineS cm}\vspace{\LineS cm}}
	\label{y_w02}
	\end{subfigure} \quad
	\begin{subfigure}{\iWidth cm}
	\includegraphics[width=  \iWidth cm ]{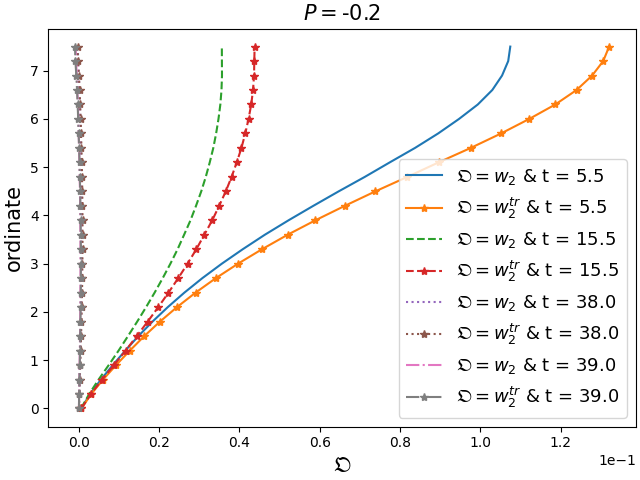}
	\centering
	\caption{Relative fluid velocity and the transformed relative fluid velocity ($\vect w^{tr} = \bar{\tens G}^{(0)} \vect w$) used in Equation \eqref{weak_form}. The comparison demonstrates the effect of the application of Nanson's formula \eqref{Nanson} (transformation of the normal vector) in mass conservation Equation \eqref{Darcy_ALE}.}
	\label{y_w_tr02}
	\end{subfigure} 
\caption{Relative fluid velocity in ALE case is calculated by $\vect w = -  \tens K (\bar{\tens F}^{(0)})^{-T} \nabla_{\vect X}p^{(0)}$ while in the linear case it is $\vect w_L = -  \tens K \nabla_{\vect x}p^{(0)}$. The comparison shows the effects of the difference in the spatial distribution of the pore pressure as well as the gradient operator transformation.}
\label{y_w}
\end{figure*}

Furthermore, the overall hydraulic response of the medium, namely, the relative fluid velocity of the ALE case is compared with the linear one in Figure \ref{y_w02} showing faster drainage and transient to steady-state transformation. An equivalent hydraulic response in reference configuration can be envisaged using the transformed relative fluid velocity shown in Figure \ref{y_w_tr02}.

Finally, for completeness of the ALE-linear comparison, we plot the micro-macro tangents, namely, the tensors $\mathbb M$ and $\tens Q$ which are also used in the calculation of the linear poroelastic model parameters. The initial values of these tensors ($M_{2222} = -0.23$, $M_{2211} = -0.076$, and $Q_{ii} = -0.113$) are used in the linear case. The significant variations in the latter components show the importance of considering different sources of nonlinearity including the nonlinear material model, the large deformation, and how their negligence could be misleading.

 \begin{figure*}
\centering
	\begin{subfigure}{\iWidth cm}
	\includegraphics[width=  \iWidth cm ]{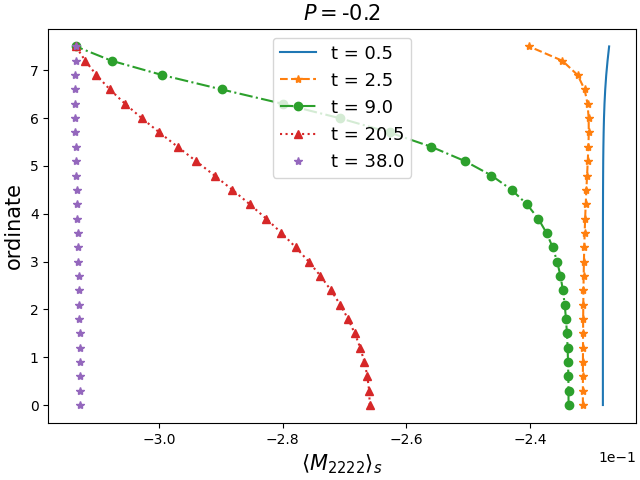}
	\centering
	\caption{The changes in diagonal element $M_{2222}$ are dominated by the negative macroscopic displacement gradient.\vspace{\LineS cm}\vspace{\LineS cm}\vspace{\LineS cm}\vspace{0.2 cm}}
	\label{y_M2222}
	\end{subfigure} \quad
	\begin{subfigure}{\iWidth cm}
	\includegraphics[width= \iWidth cm]{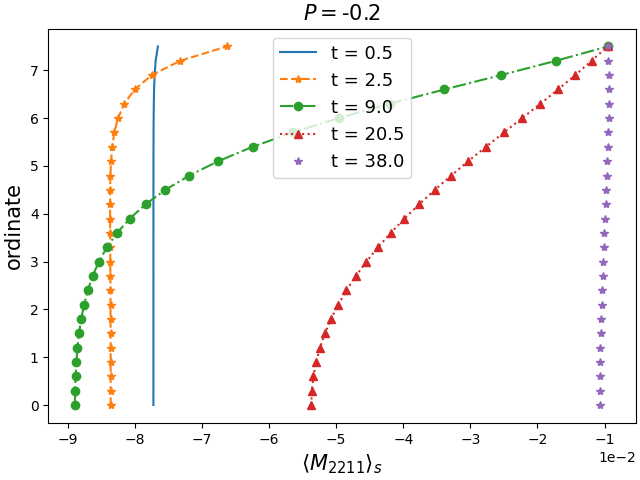}
	\centering
	\caption{The overall values of $M_{2211}$ are determined via the combination of pore pressure and $\nabla_X u^{(0)}_{22}$ such that they decrease initially at the bottom of the column (where the pore pressure is dominant) and increase approaching the top of the column and longer times.}
	\label{y_M2211}
	\end{subfigure}
	\begin{subfigure}{\iWidth cm}
	\includegraphics[width=  \iWidth cm ]{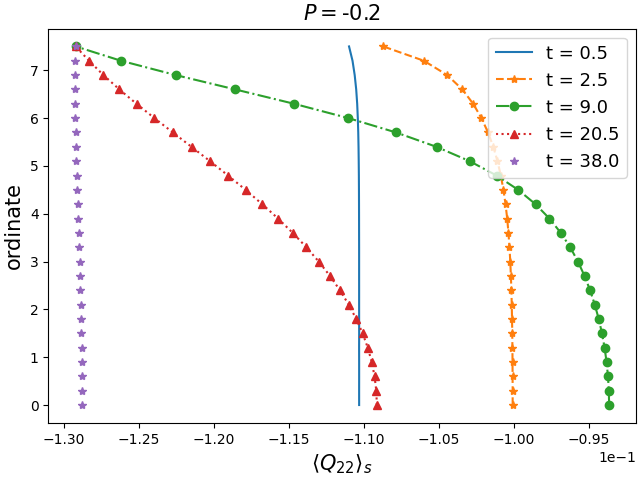}
	\centering
	\caption{The component $Q_{22}$ is affected to a large extent by both pore pressure and $\nabla_X u^{(0)}_{22}$ where the former has an increasing effect while the latter has a decreasing effect.}
	\label{y_Q22}
	\end{subfigure} \quad
	\begin{subfigure}{\iWidth cm}
	\includegraphics[width= \iWidth cm]{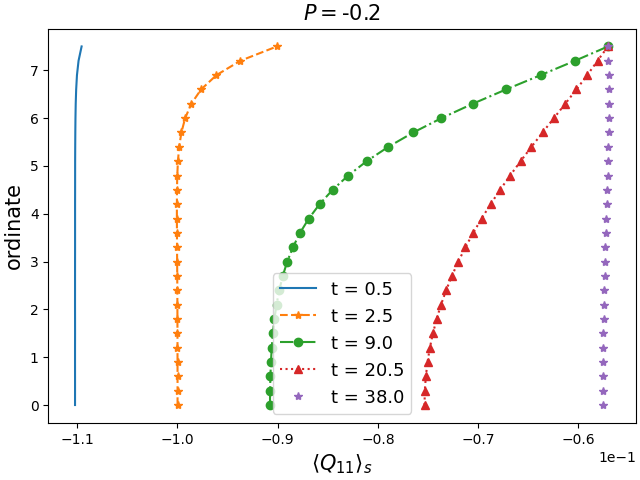}
	\centering
	\caption{Unlike $Q_{22}$, pore pressure and $\nabla_X u^{(0)}_{22}$ have increasing effects on $Q_{11}$.\vspace{\LineS cm}\vspace{\LineS cm}\vspace{0.2 cm}}
	\label{y_Q11}
	\end{subfigure}
\caption{Poroelastic properties and their variation in space and time. They are calculated by $\langle \mathbb M \rangle_s = \frac{\partial \langle \nabla_{\vect Y} \vect u^{(1)}\rangle_s}{\partial \nabla_{\vect X} \vect u^{(0)}}$ and $\langle \tens Q \rangle_s = \frac{\partial \langle \nabla_{\vect Y} \vect u^{(1)}\rangle_s}{\partial p^{(0)}}$. The initial values of these tangent tensors determine the poroelastic properties of the theory of linear poroelasticity such as effective elasticity tensor, Biot coefficient, and Biot modulus. The considerable deviation from the initial values shows the deep effects of nonlinearity in finite strain poroelasticity.}
\label{poro_prop}
\end{figure*}

\paragraph{Dimensional variables using the independent characteristic values of different scenarios of interest} 
The dimensional variables can be obtained by specifying the characteristic values of Equation \eqref{nondim2} and calculating the characterising values in Equation \eqref{non_dim} for every relevant field. For example, adopting the characteristic values of the brain tissue ($f_c = 10^{-3}[N]$, $d=2\times 10^{-6}[m]$, $\mu_c = 1[Pa\cdot s]$, and $L=10^{-3} [m]$) \cite{CSFviscosity, WANG20092371, Eskandari, KimECMsize} and of soil ($f_c = 1.5 \times 10^6[N]$, $d=2\times 10^{-4}[m]$, $\mu_c = 1[Pa\cdot s]$, and $L=1[m]$) (see also \cite{HDAZporo2021} and references therein) the dimensional variables can be obtained resembling the brain tissue via
\begin{gather}
\nonumber  \vect u_{dim}[m] = 10^{-3} \vect u,\quad \vect{w}_{dim} [mm/s] \approx 4\times 10^{-3} \vect{w}, \\ \nonumber P_{dim}[Pa]=10^3 P, \quad  t_{dim}[s] = 250 t, \\ \quad p^{(0)}_{dim}[Pa]=10^3 p^{(0)}
\end{gather}
and applied into soil mechanics
\begin{gather}
\nonumber  \vect u_{dim}[m] =  \vect u,\quad \vect{w}_{dim} [m/s]= 6\times 10^{-2} \vect{w}, \\ \nonumber P_{dim}[Pa]=1.5\times 10^6 P, \quad t_{dim}[s] = 16.6 t, \\ p^{(0)}_{dim}[Pa]=1.5\times 10^6 p^{(0)}.
\end{gather}

\section{Conclusion} \label{Conc}
Starting from the ALE formulation of the FSI problem at the pore level, we develop the governing PDEs for homogenised porohyperelastic problems under finite strain. This is accompanied by a standard fluid RVE problem (to determine the hydraulic conductivity) and a hyperelastic solid problem with pore pressure and deformation-dependent Neumann B.C. together with a body force, driven by the macroscale displacement gradient. The latter RVE problem provides, firstly, the field of microscale response to study the local phenomena and, secondly, the average microscale displacement gradient tensor. The latter is required for the homogenised governing PDEs in an online form to determine the macroscopic mechanical and hydraulic responses of the medium. Since solving the solid RVE problem via DNS for every numerical quadrature point in every time increment is very time-consuming, we employ an ANN as a real-time surrogate model for the RVE solid problem. A reliable adaptive sampling algorithm is introduced to provide an optimal training dataset for ANN parameter tuning using the Adam optimisation algorithm. 

A sensitivity analysis is carried out by varying the pore pressure and macroscale displacement gradient as the inputs of the RVE problem and the microscopic response as the outputs. A considerable deformation and strain energy density concentration is observed at the interface of the RVE pores with a magnitude much greater than the average values. We show that, considering the fully nonlinear RVE problem, several microscale deformations that are neglected in Eulerian/linear poroelastic formulation can take significant values at large deformations. Furthermore, we show that the transformed hydraulic conductivity varies nonlinearly due to the pore pressure and macroscopic deformation. 

Following the incremental weak formulation and numerical implementation of the macroscale problem (which completes the solution cycle/approach), we perform a confined compression/consolidation problem as a proof-of-concept multi-scale test for the proposed method. Although this problem results in a uniaxial macroscale displacement gradient tensor we show that the microscale deformation and, consequently, the leading order deformation gradient tensor are three-dimensional. As an implementation verification and comparison means, the same problem is also solved via linear poroelasticity using the initial effective properties derived from the same initial model parameters. As expected from the strain stiffening of the neo-Hookean material model the settlement of the nonlinear ALE case is smaller than the conventional linear poroelastic case. The importance of employing the present method is more evident by studying the hydraulic response of the medium which shows a much faster transition from transient to steady-state (faster fluid drainage). One reason for the latter is the sizeable  increase in the transformed hydraulic conductivity due to the macroscale and microscale deformations. Furthermore, the spatial variations of different parameters with their interactions are studied in detail. Finally, the model response is dimensionalised using the typical/general characteristic values in the brain tissue and soil mechanics applications showing one of the advantages of solving the problems non-dimensionally.

The present methodology is applicable in a wide range of scenarios from biological applications to soil and rock mechanics which provide numerous future directions in higher dimensions. From a numerical point of view, the solution strategy can be improved by translating the ANN into Unified Form Language (UFL) to avoid linearising it within the time increments. Furthermore, the methodology can be extended to include local fracture/damage and path-dependent RVE solid material (e.g. viscoelastic) since the response is determined using a strain energy density function.

\section*{Acknowledgement}
We acknowledge the support of this researchwork via the framework of DTU DRIVEN, funded by the Luxembourg National Research Fund (PRIDE17/12252781), and the project CDE- HUB, funded by the Luxembourg Ministry of Economy (FEDER 2018-04-024). The authors would like to thank Dr. Raimondo Penta for the invaluable discussion.

\appendix

\section{Expansion of general fields} \label{Expansion}
In this section, we apply multi-scale expansion to some basic fields, namely, $\tens F$, $J= \det(\tens F)$, and $\tens G$ plus the expanded form of Nanson's formula that are frequently used in ALE-FSI formulation.
Assuming $n=1$ which results in two-scale expansion, see Equation \eqref{power_series}, the deformation gradient may be expanded as
\begin{align}
\nonumber
\tens F &= \tens F^{(0)} + \epsilon \tens F^{(1)} \\ \nonumber
&= (\nabla_{\vect X} + \frac{1}{\epsilon} \nabla_{\vect Y})(\vect u^{(0)} + \epsilon \vect u^{(1)}) + \tens I \\ \nonumber 
&= \nabla_{\vect X} \vect u^{(0)} + \epsilon \nabla_{\vect X} \vect u^{(1)} +\frac{1}{\epsilon} \nabla_{\vect Y} u^{(0)} \\ &\hspace{3.45cm} +\nabla_{\vect Y} \vect u^{(1)} + \tens I. \label{F0expansion}
\intertext{
Considering $\mathcal C (1)$ and $\mathcal C (\epsilon^1)$ of Equation \eqref{F0expansion}, $\tens F^{(0)}$ and $\tens F^{(1)} $ are calculated as}
\tens F^{(0)} &= \nabla_{\vect X} \vect u^{(0)} + \nabla_{\vect Y} \vect u^{(1)} + \tens I \label{eqF0}\\
\tens F^{(1)} &= \nabla_{\vect X} \vect u^{(1)}.
\intertext{
We notice that considering $\mathcal C(\epsilon^{-1})$,}
0 &= \nabla_{\vect Y} \vect u^{(0)}
\intertext{
indicating that $\vect u^{(0)} $ is locally constant $\big(\vect u^{(0)}(\vect X, t)\big)$.
The Jacobian can also be expanded as}
J &= J^{(0)} + \epsilon J^{(1)} = \det(\tens F^{(0)} + \epsilon \tens F^{(1)})
\intertext{
again, considering $\mathcal C(1)$}
J^{(0)} &= \det(\tens F^{(0)}). \label{EQjzero}
\intertext{
By application of a Neumann series, $\tens F^{-1}$ can be expanded as \cite{Collis2017}}
\tens F^{-1} &= (\tens F^{(0)})^{-1} - \epsilon \tens F^{(1)} (\tens F^{(0)})^{-2},
\intertext{
which allows us to expand Piola transformation as follows} \nonumber
\tens G &= \tens G^{(0)} + \epsilon \tens G^{(1)} \\ \nonumber
&= J^{(0)} (\tens F^{(0)})^{-1} - \epsilon J^{(0)} \tens F^{(1)} (\tens F^{(0)})^{-2} \\ \nonumber
& \hspace{1cm} + \epsilon J^{(1)} (\tens F^{(0)})^{-1}  - \epsilon^2 J^{(1)} \tens F^{(1)} (\tens F^{(0)})^{-2}
\intertext{
rendering}
\tens G^{(0)} &= J^{(0)} (\tens F^{(0)})^{-1} \\
\tens G^{(1)} &= J^{(1)} (\tens F^{(0)})^{-1} - J^{(0)} \tens F^{(1)} (\tens F^{(0)})^{-2} 
\intertext{
The expansion of $\int_\Omega \nabla_{\vect X} \cdot \tens G^T = \vect 0$ results in}
\vect 0 &= \nabla_{\vect X} \cdot (\tens G^{(0)})^T + \nabla_{\vect Y} \cdot (\tens G^{(1)})^T \label{DivXG0} \\
\vect 0 &= \nabla_{\vect X} \cdot (\tens G^{(1)})^T  \label{DivXG1} \\
\vect 0 &= \nabla_{\vect Y} \cdot (\tens G^{(0)})^T.  \label{DivYG0}
\intertext{
Furthermore, we may expand Nanson's formula ($\vect n da = \tens G^T \cdot \vect N dA$), which yields} \nonumber
\vect n da &= \tens G^T \cdot \vect N {\rm d}A\\ \nonumber
&= (\tens G^{(0)T} + \epsilon \tens G^{(1)T}) \cdot \vect N {\rm d}A \\
&= \tens G^{(0)T}\cdot \vect N {\rm d}A. \label{nansonExpansion}
\end{align}

\bibliography{Dbib.bib}{}
\bibliographystyle{ieeetr} 

\end{document}